\documentclass[twocolumn]{aastex63}
\pdfoutput = 1
\usepackage{comment}
\usepackage{amsmath}
\usepackage{mathrsfs}
\usepackage{graphicx}

\usepackage{multirow}
\usepackage{graphicx}
\usepackage{color}

\newcommand\aastex{AAS\TeX}
\newcommand\latex{La\TeX}

\begin{document}

\title{Unveiling the Interior Structure and Thermal Evolution of Super-Earth GJ 486b}

\correspondingauthor{Liton Majumdar}
\email{liton@niser.ac.in, dr.liton.majumdar@gmail.com}

\author[0000-0003-1057-2320]{Chandan K. Sahu}
\affiliation{Exoplanets and Planetary Formation Group, School of Earth and Planetary Sciences, National Institute of Science Education and Research, Jatni 752050, Odisha, India}
\affiliation{Homi Bhabha National Institute, Training School Complex, Anushaktinagar, Mumbai 400094, India}

\author [0000-0001-7031-8039] {Liton Majumdar}
\affiliation{Exoplanets and Planetary Formation Group, School of Earth and Planetary Sciences, National Institute of Science Education and Research, Jatni 752050, Odisha, India}
\affiliation{Homi Bhabha National Institute, Training School Complex, Anushaktinagar, Mumbai 400094, India}

\author [0009-0001-0265-7095] {Sudipta Mridha}
\affiliation{Exoplanets and Planetary Formation Group, School of Earth and Planetary Sciences, National Institute of Science Education and Research, Jatni 752050, Odisha, India}
\affiliation{Homi Bhabha National Institute, Training School Complex, Anushaktinagar, Mumbai 400094, India}
\affiliation{Department of Physical Sciences, Indian Institute of Science Education and Research, Kolkata, Mohanpur 741246, West Bengal, India }

\author [0009-0000-2964-9450] {Harshit Krishna}
\affiliation{Exoplanets and Planetary Formation Group, School of Earth and Planetary Sciences, National Institute of Science Education and Research, Jatni 752050, Odisha, India}
\affiliation{Homi Bhabha National Institute, Training School Complex, Anushaktinagar, Mumbai 400094, India}
\affiliation{Department of Physical Sciences, Indian Institute of Science Education and Research (IISER) Mohali,
Knowledge City, Sector 81, Sahibzada Ajit Singh Nagar, Punjab 140306, India}

\begin{abstract}

Recent ground- and space-based surveys have shown that planets between Earth and Neptune in size, known as “super-Earths,” are among the most frequently found planets in the Galaxy. Although the JWST era has provided high-quality atmospheric data on many such super-Earths, modeling tools are crucial for understanding their unobservable interiors. Consequently, interior studies represent the next essential step in gaining a comprehensive understanding of this class of exoplanets. This study investigates the interior structure, thermal evolution, and atmospheric dynamics of the super-Earth GJ 486b using \texttt{SERPINT}, a 1-D self-consistent coupled interior structure and evolution model, aiming to understand the planet's thermal evolution based on an Earth-like structure. Our results indicate that GJ 486b's core is approximately 1.34 times larger than Earth's, with a core pressure of about 1171 GPa. The thermal evolution model predicts that the planet's mantle cools and solidifies over approximately 0.93 million years. As the magma ocean cools, water is released from the melt, forming a water-rich atmosphere during early solidification. Photolysis of water vapor and subsequent hydrogen escape lead to oxygen accumulation, forming a water- and oxygen-rich secondary atmosphere. Future high-sensitivity JWST observations, with improved wavelength coverage and the detection of additional trace gases, will enable a detailed analysis of the planet's atmospheric composition, providing crucial insights into the interior, surface, and subsurface properties of GJ 486b.

\end{abstract}


\section{Introduction} \label{sec:intro}

Since the discovery of the first rocky exoplanet, Kepler-10b \citep{batalha2011kepler}, numerous terrestrial exoplanets have been identified, highlighting the existence of Earth-sized planets revolving around their host stars. The TRAPPIST-1 system is one of the few exoplanetary systems discovered that consists of seven terrestrial exoplanets, three of which lie within the habitable zone of their host star \citep{luger2017seven, grimm2018nature, turbet2020review}. Recent work indicates that TRAPPIST-1b has an atmosphere-less surface \citep{de2016combined}. \citet{zieba2023no} ruled out the presence of a thick atmosphere on TRAPPIST-1c and found evidence of a thin CO$_2$ atmosphere. 
Thanks to the wealth of rocky exoplanet discoveries and advances in observing capabilities, the effort to characterize rocky terrestrial exoplanets has experienced significant growth within the scientific community \citep{2022ARA&A..60..159W}. Analogous to Earth, terrestrial exoplanets provide insights into diverse conditions and potential habitability that might exist beyond Earth. These planets are typically characterized by Earth-like densities and thin atmospheres, with interiors predominantly composed of rock, silicate melts, and/or liquids (such as water oceans) \citep{VanHoolst2019}. Their atmospheres typically exhibit radii less than $\approx 1.5$ Earth radii ($R_{\oplus}$) and comprise a thin layer of volatiles, resulting in secondary atmospheres formed by outgassing from magma oceans, vaporization of volatiles, and volcanic activities \citep{lopez2014understanding, Unterborn2019}.

There are several different types of atmospheres that can form on terrestrial planets through various mechanisms. Some of these include the formation of secondary atmospheres by volcanic outgassing \citep{schaefer2016predictions, ortenzi2020mantle, krissansen2022predictions, liggins2022growth}, steam atmospheres resulting from ice pebbles accumulating around the water ice line \citep{mousis2019jupiter, krissansen2021oxygen, kimura2022predicted}, and the formation of silicate atmospheres \citep{schaefer2009chemistry, schaefer2012vaporization, zilinskas2022observability}. These diverse atmospheric compositions are influenced by various interior compositions, introducing degeneracies in the internal structure; for instance, planets of identical mass and radius can differ in interior structure and mineralogy \citep{seager2007mass, huang2022magrathea}.
A planet with a dense iron core and a light mantle may have the same mass and radius as a planet with a relatively light core and a dense mantle. Both types of planets, based on their mineral compositions, may produce very similar or completely different atmospheres \citep{madhusudhan2012c, dorn2017generalized, Unterborn2019}. 

During the slow cooling phase of the magma ocean stage, the structure of the core and mantle develops as minerals and rocks differentiate based on density gradients \citep{rubie2011heterogeneous, Dorn2018, schaefer2018magma, tronnes2019core}. The initially homogeneous magma separates, with heavier iron content sinking to the bottom, resulting in the formation of the metallic core \citep{brett1984chemical, righter2003metal, litasov2016composition}, quite analogous to small bodies like Ceres, which may not have a fully differentiated interior due to a short magma ocean stage due to which the iron did not entirely separate \citep{zolotov2009composition, neveu2015geochemistry, zolotov2020composition}. Understanding composition, density gradients, and core characteristics informs mass-radius relations and planetary classification, aiding predictions about atmospheric composition, volatile retention, and geological activity, which impact habitability assessments \citep{schaefer2018magma, VanHoolst2019, nettelmann2021exoplanetary}.



Terrestrial planets have diverse surfaces with solids and liquids vital for complex chemical reactions dependent on surface activation, which atmospheric processes alone might hinder \citep{berdyugina2019surface}.
Additionally, the composition and dynamics of a planet's interior are pivotal in assessing these reactions. Earlier work by \citet{dorn2021hidden} indicates that water in the deep mantle and core may contribute up to 16\% uncertainty in the planet's bulk radius. The interaction between the atmosphere of a terrestrial planet and its interior is crucial, as its gravity may not be strong enough to retain primary atmospheres composed of H$_2$-He during the early accretion \citep{kite2016atmosphere, kite2020exoplanet}. Consequently, these planets are unlikely to have a primary H$_2$-He atmosphere due to insufficient gravitational pull, resulting in secondary atmospheres with higher molecular masses \citep{elkins2008ranges, suer2023distribution}.
Exoplanets such as 55 Cnc e \citep{demory2016variability, crida2018mass, dorn2019new, hu2024secondary} and GJ 1132b \citep{schaefer2016predictions, diamond2018ground, may2023double} have been extensively studied for their rocky interiors and high molecular mass secondary atmospheres.


The unobservability of the interiors of rocky exoplanets highlights the importance of theoretical and numerical models to understand their interior structure and evolution. 
While several models exist to study planetary interiors, recent models such as \texttt{ExoPlex} \citep{lorenzo2018exoplex, Unterborn2019} and \texttt{Magrathea} \citep{huang2022magrathea} include different interior compositions to simulate the structure of the planet. In contrast, thermal evolution models such as \texttt{VPLANET} \citep{barnes2020vplanet, barth2021magma} and \texttt{PACMAN} \citep{krissansen2021oxygen, krissansen2022predictions} incorporate parameterized models in conjunction with the atmosphere. Each model operates in its own region of parameter space and incorporates slightly different physics. Studying terrestrial exoplanets involves comparing them with rocky planets in our solar system, including moons and dwarf planets \citep{greenwood2005widespread, elkins2008linked, hosono2019terrestrial, salvador2023magma}. These comparisons help draw parallels and discern commonalities, refining our understanding of planetary system dynamics. Examining constraints like surface composition and atmospheric dynamics provides deeper insights into planetary formation and evolution \citep{turrini2015role}.

While previous models have explored the structure and thermal evolution of rocky interiors, a well-defined coupling between these two aspects and the formation of the atmosphere requires further investigation. This work examines the interior structure, thermal evolution, and atmospheric development of the rocky planet GJ 486b, which has a mass of 3.00 M$_{\oplus}$ and a radius of 1.343 R$_{\oplus}$, orbiting an M-dwarf star \citep{caballero2022detailed}.

GJ 486b orbits a bright, weakly active M-dwarf host star with a short orbital period of 1.47 days and a warm surface temperature, making it a prime target for atmospheric characterization through emission and transit spectroscopy with JWST \citep{caballero2022detailed}. For instance, it was recently observed using high-precision photometry and spectroscopy techniques with JWST's NIRSpec/G395H instrument. These observations have provided valuable insights into its structure and atmospheric composition \citep{caballero2022detailed, ridden2023high, moran2023high}. Specifically, two transits of GJ 486b were observed using NIRSpec in the G395H grating mode, covering wavelengths from 2.87 to 5.14 $\mu$m at an average native spectral resolution of $R \sim 2700$ \citep{moran2023high}. This dataset was used to analyze the transmission spectrum of GJ 486b and investigate the presence of atmospheric features, particularly water vapor. The data, reduced independently using three pipelines (Eureka! \citep{bell2022eureka}, FIREFLy \citep{rustamkulov2022analysis, rustamkulov2023early}, and Tiberius \citep{kirk2018lrg, kirk2019lrg, kirk2021access}), consistently revealed a non-flat spectrum with a slight blueward slope ($\leq 3.7$ $\mu$m), challenging the notion of a flat spectrum with a confidence level between 2.2$\sigma$ and 3.3$\sigma$. This could be interpreted as evidence for either a water-rich atmosphere or stellar contamination from unocculted cool spots on the host star \citep{moran2023high}.

Retrieval analyses using the POSEIDON code \citep{macdonald2017hd, macdonald2023poseidon} favor the presence of a water-dominated atmosphere with a mixing ratio greater than 10\% at 2$\sigma$ confidence for the Eureka! reduction \citep{moran2023high}. At the same time, the possibility of CO$_2$ was also observed to deviate from the fitting \citep{moran2023high}.
The observed spectral slope of GJ 486b could also be due to cool starspots on the host star, with a spot coverage fraction of about 10\%. This suggests that the transmission spectrum is influenced by H$_2$O absorption in these spots rather than the planet's atmosphere \citep{moran2023high}. \color{black}
However, this JWST dataset cannot definitively rule out either of these scenarios.


The structure of this paper is as follows: Section \ref{sec:methods} describes our model incorporated to understand the interior of rocky planets. In Sections \ref{subsec:structure} and \ref{subsec:composition}, we describe the structural model, detailing the equations of state and interior physics used in the model. Section \ref{subsec:thermal} describes the thermal model, which is coupled with the atmospheric model discussed in Section \ref{subsec:atmosphere}. We then present our results for GJ 486b in Section \ref{sec:results} and discuss our findings in Section \ref{sec:discussion}. Finally, we summarize our results in Section \ref{sec:conclusion} and conclude with our prospects.

\section{Methods} \label{sec:methods}

\begin{figure*}[ht]
    \centering
    \includegraphics[width=\textwidth]{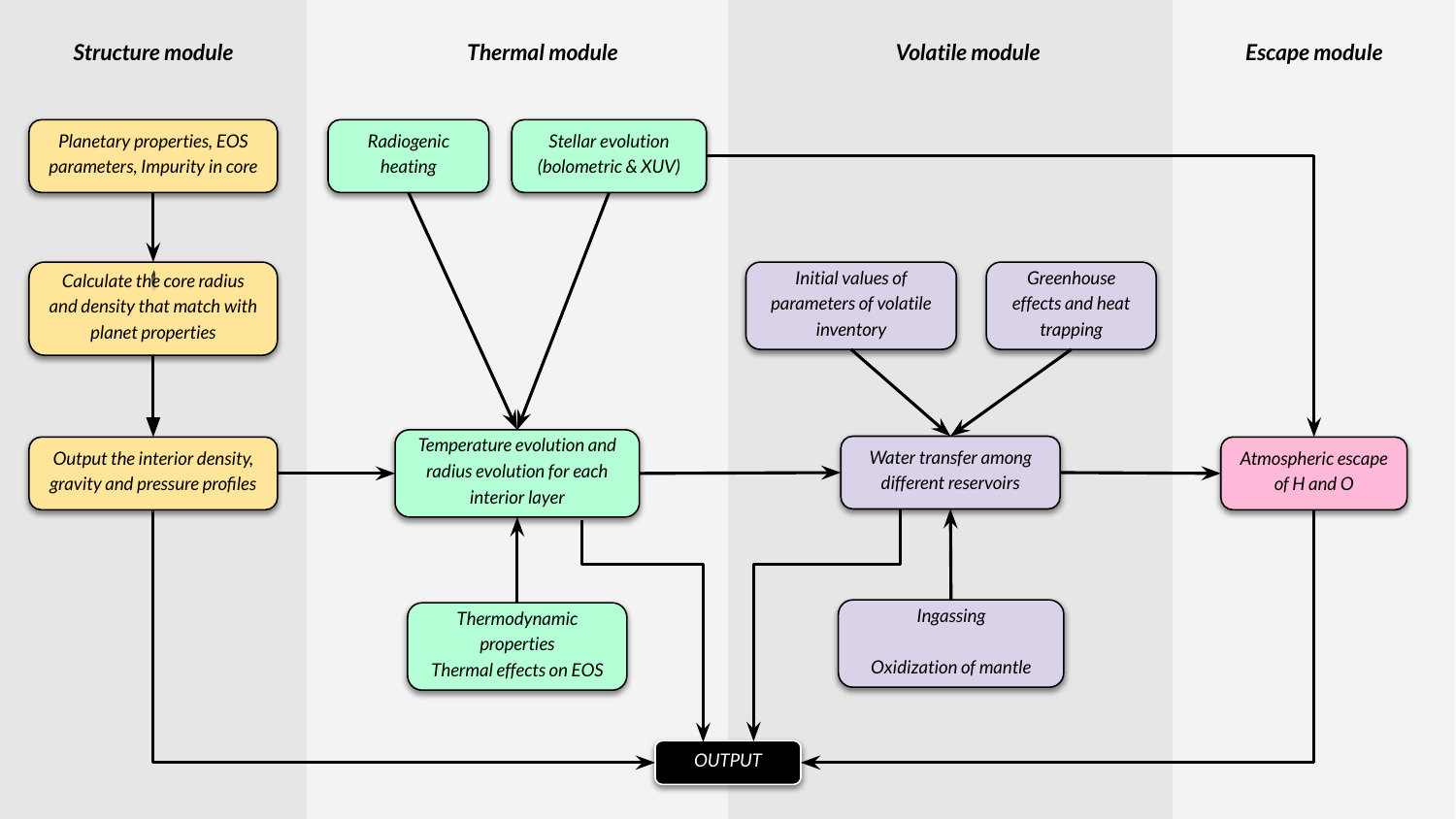}
    \caption{Flowchart of \texttt{SERPINT} illustrating the four different modules of the model: structure, thermal, volatile, and escape  (arranged from left to right), along with the interactions among them.}
    \label{Flowchart}
\end{figure*}

We introduce \texttt{SERPINT} (Structure and Evolution model for Rocky Planet INTeriors), a 1-D spherical model designed to simulate both the interior structure and the coupled thermal-atmospheric evolution of rocky exoplanets around M-dwarfs. In this section, we present the various components of \texttt{SERPINT}, summarized in Fig. \ref{Flowchart}. 

Assuming an Earth-like composition with a fully differentiated core containing silicon, sulfur, and oxygen impurities, and a mantle composition similar to that of Earth, we modeled the interior structure of GJ 486b in analogy to Earth, which is the most extensively studied rocky planet. For the secondary atmosphere of GJ 486b, we assume a pure water composition, based on \citet{moran2023high}. We begin this section by explaining the model parameters and the governing equations for the interior structure of GJ 486b, followed by a detailed discussion of its thermal and atmospheric evolution.

\subsection{Structure of the interior} \label{subsec:structure}

In this section, we describe the interior model and its various constituents. 
Seismological data suggests that the Earth has a two-layered Fe-Ni core and a thick mantle comprising 84\%  of its volume \citep{dziewonski1981preliminary}. 
The Earth’s inner core is solid, whereas the outer core is predominantly liquid. 
In contrast, the mantle is primarily solid. 

We used equations of states (EOS) for the core, mantle, and crust to model the relation between each layer and the geological evolution of terrestrial planets.  
We use the standard approach to parameterize the planet's interior structure based on the densities of interior layers and solve the hydrostatic equilibrium equations.
The mantle is set up to be in the form of a magma ocean that evolves to give the interior structure of the planet.

The gravitational acceleration of a planet can be parameterized by the density profile using the hydrostatic equilibrium equation
\begin{equation}
    g = \int_0^{r} \left(\frac{4}{3}\pi G r'\dfrac{d\rho}{dr'} + 4 \pi G  \rho\right)dr'
    \label{eqn:grav}
\end{equation}
here $g$ is the gravitational acceleration of the planet at radius $r$ bounded by materials with density $\rho(r)$. $G$ is the gravitational constant.
The interior pressure profile can be related to density and gravity as 
\begin{equation}
    P = P_0 + \int_{\delta}^{R_{\text{p}}} g \rho d\delta
    \label{eqn:pres}
\end{equation}
where $P_0$ is the atmospheric pressure, $\delta$ is the depth from the planet's surface and $R_p$ is the radius of the planet.
The planet structure is obtained by solving the equations of hydrostatic equilibrium numerically, where $m$ is the mass enclosed within radius $r$.
\begin{align}
    \dfrac{dm}{dr} = 4 \pi r^2 \rho
\end{align}
\begin{align}
    \dfrac{dP}{dr} = -\dfrac{G \rho m}{r^2}
\end{align}

We solve these equations using different EOSs for different layers of the interior, which are described in detail in section \ref{subsec:composition}.

\subsection{Interior composition} \label{subsec:composition}

Works by \citet{Pekmezci_2020, spaargaren2023plausible} show refractory materials present in most protoplanetary disks are comparable to Earth's internal composition.
This implies that the interior composition of terrestrial planets formed from these disks can be assumed to be similar to Earth's.
However, being around diverse kinds of stars and protoplanetary disks, these planets may contain other elements too \citep{hinkel2024host, helling2019exoplanet}, and hence can give rise to variations in the interior structure.
Here, we adopt a simple approach to model the exoplanetary interior using Earth's composition. 
The planet's interior in our model comprises the core and a two-layered mantle (upper and lower mantle) \citep{seager2007mass, Unterborn2019}. 
However, contrary to \citet{seager2007mass, Unterborn2019}, we have not included any water layer in our structure model. The core is believed to be composed mainly of iron, with small amounts of nickel and lighter elements, such as sulfur and oxygen. 
We have excluded the crust in this model, as the Earth's crust is the product of numerous complex processes that are beyond the scope of this work.

\subsubsection{The core} \label{subsubsec:core}

The majority of Earth's core is in a liquid state, with approximately 5\% of core mass consisting of solid material \citep{dziewonski1981preliminary, yoder1995astrometric}.
However, studies by \citet{unterborn2016scaling} indicate that the state of the core (solid/liquid), amount of light elements in the core and the composition of the mantle can produce uncertainties in the core radius.
While modeling liquid-solid state transitions is feasible, studies like \citet{valencia2006internal, valencia2007radius} have demonstrated the complexity and limitation of simulating geotherms for other planets. 
Moreover, they highlighted the potential ubiquity of liquid cores in super-Earth-sized planets.
Therefore, assuming a 100\%  liquid core for a basic model like ours is empirically more consistent.
A deficiency in density within the Earth's core compared to pure iron suggests the presence of light elements, including silicon, nickel, sulfur, and oxygen \citep{birch1964density, anderson1994equation, poirier1994light}. 
These uncertainties in the state and composition of the core must be addressed properly when modeling exoplanets in other planetary systems.
Our model employs a mineralogical system consisting of pure Fe, FeS, FeO, and Fe$_2$O$_3$.
We have used a simplified core-mantle boundary rather than a fully diffused model.

Several EOSs may be utilized to study the pressure-density relations of the interior. 
Some of the most common EOSs include
\begin{enumerate}
    \item Third order Birch-Murnaghan (BME3) EOS \citep{birch1964density, birch1988elasticity,seager2007mass}
    \begin{equation}
        \begin{split}
        P = &\dfrac{3K_0}{2} \left(\xi^{7/3} - \xi^{5/3} \right) \\
        & \times \left[ 1+\dfrac{3}{4}\left(K'_0-4\right)\left(\xi^{2/3}-1\right) \right]
        \label{eqn:BME3}
    \end{split}
    \end{equation}

    \item Vinet EOS \citep{vinet1986universal, vinet1989universal, seager2007mass}
    \begin{equation}
    \begin{split}
        P = &3K_0 \xi ^{2/3}\left(1 - \xi^{-1/3} \right) \\
        & \times \exp \left[\dfrac{3}{2}(K_0'-1) \left(1 - \xi^{-1/3} \right)\right]
        \label{eqn:vinet}
    \end{split}
    \end{equation}

    \item Keane EOS \citep{keane1954investigation, zhang2022thermal}
    \begin{equation}
    \begin{split}
        P = & \frac{K_0 K_0^{\prime}}{K_{\infty}^{\prime 2}}\left(\xi^{K_{\infty}^{\prime}}-1\right) - \left(\frac{K_0 K_0^{\prime}}{K_{\infty}^{\prime}} - K_0 \right) \ln \xi
        \label{eqn:keane}
    \end{split}
    \end{equation}
    
    \item Holzapfel EOS \citep{holzapfel1996physics, holzapfel1998equations}
    \begin{equation}
    \begin{split}
        P = P_0 \:+\: &3 K_{0} \xi^{-5/3}\left[1+c_2 \xi^{1/3}\left(1-\xi^{1/3}\right)\right] \\
        & \times \left[1-\xi^{1/3}\right] \exp \left[c_0\left(1-\xi^{1/3}\right)\right]
        \label{eqn:holzapfel}
    \end{split}
    \end{equation}
    where $\rho_0 = \mu_{\text{Fe}}$/$V_0$, $V_0 = 4.28575$ cm$^3$/mol, $P_0 = 234.4$ GPa, $K_0 = 1145.7$ GPa, $c_0 = 3.19$ and $c_2 = -2.40$  \citep{hakim2018new}. 
\end{enumerate}

Here $\rho_0$ is the reference density, $\xi = \rho/\rho_0$, $K_0 = -V(\partial P/\partial V )_T$ is the isothermal bulk modulus, $K_0'$ is the first order derivative of bulk modulus defined at the reference temperature and zero pressure, and $K_{\infty}^{\prime}$ is the derivative of bulk modulus in the infinite pressure limit. 

It is still debatable which EOS governs the interior properties for Earth best \citep{wagner2011interior}. 
The \texttt{SERPINT} model defines the core as being composed primarily of $\varepsilon$-Fe liquid, i.e., the high-pressure phase of iron with a hexagonal close-packed (HCP) crystal structure called hexaferrum. 
We utilize the Holzapfel equation of state (Eqn. \ref{eqn:holzapfel}) for its fewer volume residuals for pressures above 234 GPa compared to Vinet and BME3 EOS for the iron core \citep{hakim2018new}.
The calculated EOS for liquid $\varepsilon$-Fe suggests a good fit over the Earth's core.
We added an impurity factor ($f$) to the core density assuming S, O and Si impurities by weight, which reduce the density ($\rho_0$) to $\rho_0(1-f)$.
This parameter is used to modify the core's density and provide a better empirical fit to the Earth's core. 
(Refer Appendix \ref{apdx:sec:impurity} for more information.)





\subsubsection{The mantle} \label{subsubsec:mantle}

The mantle surrounds the core and is primarily composed of silicate minerals, such as olivine, pyroxene, garnet, perovskite (Pv), post-perovskites (pPv), and others.
The Earth's lower mantle, characterized by high pressure and temperature, contains silicate minerals rich in magnesium and iron \citep{Hemley2013}. 
Convection currents during the magma ocean phase in this region drive the movement of molten rocks, affecting heat distribution, magma flow, and volatile transport \citep{o1977scale}.
Above the lower mantle, the upper mantle extends toward the planet's surface, where silicate minerals with lower density and less compact structures reside. The asthenosphere, a ductile, partially molten layer of the upper mantle just below the lithosphere, allows the rigid lithospheric plates to move. Together, these layers facilitate plate tectonics \citep{anderson1995lithosphere}.


We adopt a simple two-layered mantle structure in \texttt{SERPINT}, comprising low-density silicates like olivine in the upper mantle and higher-density peridotites such as bridgmanite, pyroxene, perovskites, and post-perovskites in the lower mantle. 
We employ the Vinet EOS (Eqn. \ref{eqn:vinet}) in the upper mantle, which provides a better fit for the Preliminary Reference Earth Model (PREM) than the third-order Birch-Murnaghan (BME3) EOS.
We have used $\rho_0 = 4105.9$ kg m$^{-3}$, $K_0 = 267.7$ GPa, and $K_0' = 4.04$ GPa for the lower mantle and $\rho_0 = 3213.7$ kg m$^{-3}$, $K_0 = 125.0$ GPa, and $K_0' = 5.00$ GPa for the upper mantle in our model \citep{seager2007mass}. 
A pressure limit of 23.5 GPa separates the upper mantle from the lower mantle, which also denotes the olivine-perovskite (Ol-Pv) transition \citep{hirschmann2000mantle, fei2004experimentally}, while pressure limit of 125 GPa denotes the perovskite-post-perovskite (Pv-pPv) transition \citep{murakami2004post, hirose2006postperovskite}.  
At the beginning of our model, we assume the planet to be an entirely molten homogeneous mixture of its constituents. The density profile changes as the planet cools down, and we get a differentiated planet interior.

\subsection{Thermal model} \label{subsec:thermal}

This section describes the thermal evolution model in \texttt{SERPINT}. 
Terrestrial planets usually have very high temperatures at the time of their formation due to the heat of accretion, leading to partial or complete melting, resulting in the formation of magma oceans, which is also the starting point of our model \citep{chambers1998making, morbidelli2012building}. 
The planet gradually undergoes cooling, accompanied by exchanges between reservoirs.
Our thermal model is adapted from \citet{schaefer2016predictions} and \citet{barth2021magma}.

\subsubsection{Stellar heating} \label{subsubsec:stellar}

During planetary evolution, the host stars undergo significant evolution over large timescales. 
Most of the rocky planets discovered so far lie around M-dwarf stars \citep{shields2016habitability}.
These are red low-mass stars with a mass between 0.08 to 0.6 $M_\odot$ \citep{henry2006solar, henry2018solar}, and commonly emit X-rays and UV radiation, with much smaller XUV luminosities \citep{adams2005m, baraffe2015new}.
The XUV luminosity is about $10^3$ times smaller than bolometric luminosity for solar-like stars after $10^5$ years \citep{luger2015extreme}.
Due to slow fuel consumption, M-dwarf stars can maintain nearly constant luminosity for about a few million years. 
Their luminosity evolution is characterized by saturation timescale $t_{sat}$, saturation flux $f_{sat}$, and the exponential reduction factor $\beta_{XUV}$.
\texttt{VPLANET} \citep{barnes2020vplanet} based on the \citet{baraffe2015new} stellar model uses $f_{sat}=10^{-3}$, $t_{sat}=10^9$ years, and $\beta$ = 1.23. \texttt{PACMAN} \citep{krissansen2022predictions} samples these parameter values randomly from a small range.
The model by \citet{baraffe2015new} consists of a grid of stellar evolution data with masses ranging from 0.1 $M_{\odot}$ to 1.4 $M_{\odot}$. 
They calculated the variation of the XUV radiation of M-dwarfs around their mean lifetimes of $\approx 10^{10}$ years. 
The details of the stellar grid model can be found in \citet{baraffe1998evolutionary}.
We use the same grid of models for our stellar heating part after extrapolating the data for the initial ages of the star assuming stable illumination.
The XUV luminosity for the host star is given by \citep{luger2015extreme}
\begin{equation}
\dfrac{L_{XUV}}{L_{bol}} = 
    \begin{cases}
        f_{sat}, \qquad \qquad \qquad &  t\leq t_{sat} \\ \\
        f_{sat} \left( \dfrac{t}{t_{sat}} \right) ^{-\beta}, &  t > t_{sat}
    \end{cases}
    \label{eqn:XUVlum}
\end{equation}
Stars with $t_{sat}$ in the 7 to 9 billion years range were sampled from \citet{birky2021improved} as it provides the best posterior distributions.
To determine the star’s age at the outset, we reference the estimated lifespan of the protoplanetary disk.
Based on observations of the solar neighbourhood, \citet{ribas2014disk} reported that low-mass star disks typically persist for 4.2 to 5.8 million years. 
Therefore, we adopt the host star's initial age of 5 million years in all simulation scenarios.

\subsubsection{Radiogenic heating} \label{subsubsec:radiogenic}

\begin{figure*}[ht]
    \centering
    \includegraphics[width=0.45\textwidth]{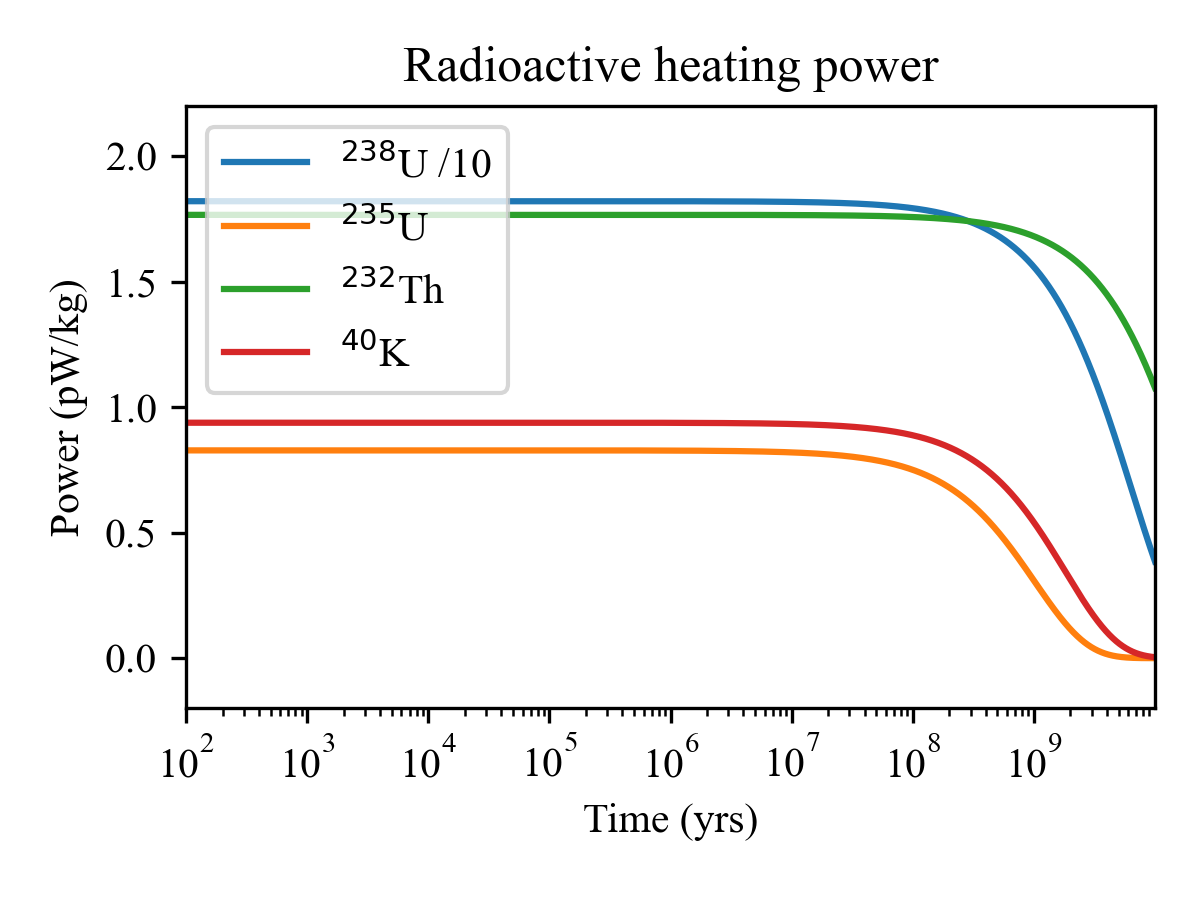}
    \caption{Evolution of radiogenic heating power per unit mass for the radionuclides $^{238}$U, $^{238}$U, $^{238}$Th and $^{40}$K} (the heating power of $^{238}$U have been factored by 10.)
    \label{fig:rad_heat}
\end{figure*}

A large portion of the Earth's heat originates from the radioactive decay of elements inside its interior, which prolongs the magma ocean phase \citep{Plant1996}.
These include short- and long-lived radionuclides such as $^{238}$U, $^{235}$U, $^{232}$Th, $^{40}$K, and $^{26}$Al. 
Our radioactive heating scheme is similar to \citet{lebrun2013thermal}. 
The radioactive decay equation can be written as 
\begin{align}
    Q = \sum_i Q_{0, i} \text{ exp} \left( -\dfrac{\text{ln(2)}\:t}{\tau_i} \right)
    \label{eqn:Qr}
\end{align}
where $Q_{0, i}$ and $\tau_i$ are the initial heat production rate per unit mass and the half-life of the radioactive species $i$ at t = 0, respectively. 
For individual radioactive species $i$, we can write $Q_{0, i}$ as
\begin{equation}
    Q_{0, i} = N_{i} \times c_{i} \times H_{i}
\end{equation}
where $N_{i}$ is the natural abundance, $c_{i}$ is the current concentration and $H_{i}$ is the heat production per unit mass of the radioactive species $i$.
The radio-nuclides used in our model include $^{238}$U, $^{235}$U, $^{232}$Th, and $^{40}$K. 
We do not use $^{26}$Al due to its comparatively short half life and uncertainty in the concentration and time of deposition in the history \citep{lebrun2013thermal, ruedas2017radioactive}. 
Fig. \ref{fig:rad_heat} shows the evolution of heating power produced by the four radionuclides per unit mass for Earth-like radionuclide concentrations.
A complete description of the mentioned radionuclide half-life, abundance, and concentration are taken from \citet{lebrun2013thermal}.

\subsubsection{Mantle cooling and solidification} \label{subsubsec:solidification}

The thermal model integrates the atmosphere, interior, and host star dynamics for analyzing Earth-like exoplanets.
Simulations begin with a molten mantle, gradually solidifying from the bottom up.
The magma ocean cools by convecting molten rocks towards the surface, eventually dissipating the heat from the interior to space through radiation. 
Meanwhile, stellar radiation heats the surface, with some of it being reflected into space because of the albedo of the planet.

\begin{figure*}[ht]
    \centering
    \includegraphics[width=0.45\textwidth]{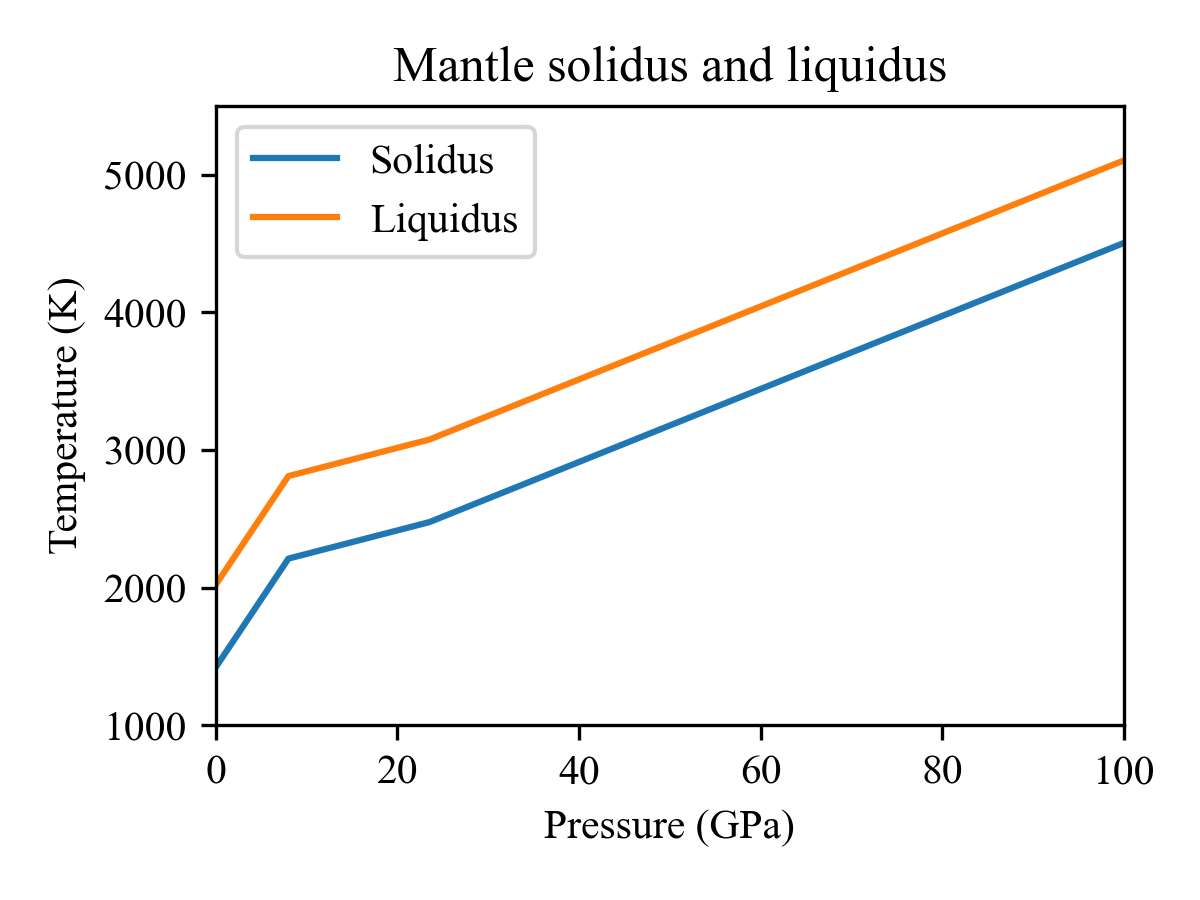}
    \caption{Solidus and liquidus profiles used in \texttt{SERPINT}}
    \label{fig:T_sol_liq}
\end{figure*}

The heating and cooling of the planet is monitored through  the potential temperature ($T_{\text{p}}$), the surface temperature ($T_{\text{surf}}$), and the solidification radius ($\text{r}_\text{s}$) at each time step. 
Mantle potential temperature ($T_{\text{p}}$) is the temperature that mantle material would have if it were brought to the surface adiabatically.
Given the high efficiency of convection within the molten mantle, thermal equilibrium is maintained between the magma ocean and the surface, validating the assumption of $T_{\text{surf}} = T_{\text{p}}$. 
Additionally, the model governs the evolution of melt fraction ($\phi$), solidification radius ($r_{\text{s}}$), and mantle viscosity ($\eta$) \citep{abe1997thermal, lebrun2013thermal}. 
At the beginning of the simulation, $T_{\text{p}}$ and $r_{\text{s}}$ are initialized at 5000 K to begin with a largely molten mantle.
The adiabatic temperature structure of the mantle, $T(r)$, is a function of planet radius as described in \citet{schaefer2016predictions}, which can be related to the mantle potential temperature $T_{\text{p}}$ as 
\begin{align} 
    T \left( r \right) = T_{\text{p}} \left( 1 + \dfrac{\alpha g_{\text{s}}}{C_{\text{p}}} \left( R_{\text{p}} -r \right) \right)
    \label{eqn:T_eqn}
\end{align}
where $\alpha$ is the thermal expansion coefficient, $g_{\text{s}}$ is the planetary surface gravity, $R_{\text{p}}$ is the radius of the planet and $C_{\text{p}}$ is the specific heat capacity of the magma ocean.
To parameterize the solidification, we use the solidus-liquidus profiles from \citet{hirschmann2000mantle, hirschmann2009dehydration}, similar to \citet{barth2021magma, krissansen2022predictions}.
We again simplify this curve into linear form using the prescription shown in Eqn. \ref{eqn:T_solidus}, which has also been shown in Fig. \ref{fig:T_sol_liq}
\begin{equation}
T_{\text{solidus}}(P) = 
    \begin{cases}
        98.8 \times 10^{-9} P + 1420, \qquad & P\leq 8 \\
        17.07 \times 10^{-9} P + 2073.86, \qquad & 8 \leq P \leq 23.5 \\
        26.53 \times 10^{-9} P + 1851.55, \qquad & 23.5 \leq P 
    \end{cases}
    \label{eqn:T_solidus}
\end{equation}

Here $P=\rho g_{\text{s}} (R_{\text{p}} - r)$ is in GPa.
Upon combining both equations and parameterizing $r_{\text{s}}$, we get
\begin{align}
    r_{\text{s}}=R_{\text{p}}-\dfrac{T_{\text{p}}-b}{a g_{\text{s}} \rho_{\text{m}}-\dfrac{\alpha g_{\text{s}}}{C_{\text{p}}}\,T_{\text{p}}}
    \label{eqn:r_s}
\end{align}
\begin{align}
    \dfrac{dr_{\text{s}}}{dt}=\dfrac{C_{\text{p}}}{g_{\text{s}}}\dfrac{\left(b\alpha-a\rho_{\text{m}}C_{\text{p}}\right)}{\left(a\rho_{\text{m}}C_{\text{p}}-\alpha T_{\text{p}}\right)^{2}}\dfrac{dT_{\text{p}}}{dt}
    \label{eqn:dr_s_dt}
\end{align}

\citet{schaefer2016predictions} describe the magma ocean temperature as
\begin{align}
    V_{\text{m}}C_{\text{p}}\dfrac{dT_{\text{p}}}{dt}\, = & \,4\pi r_{\text{s}}^{2}\rho_{\text{m}}\dfrac{dr_{\text{s}}}{dt}\Delta H_{\text{f}}\,-\,4\pi R_{\text{p}}^{2}F\,+\,V_{\text{m}}Q_{\text{r}}
    \label{eqn:dT_p_dt}
\end{align}
where $V_{\text{m}} = \dfrac{4\pi}{3}\rho_{\text{m}}\left(R_{\text{p}}^{3}-r_{\text{c}}^{3}\right)$ is the volume occupied by the mantle, $\rho_{\text{m}}$ bulk density of the mantle, $C_{\text{p}}$ is the silicate heat capacity ($1.2 \times 10^3$ J kg$^{-1}$ K$^{-1}$), $\Delta H_{\text{f}}$ is the heat generated by the fusion of silicates ($4 \times 10^5$ J kg$^{-1}$), $F$ is the heat flux from the mantle and $Q_{\text{r}}$ is the heat originating from the radiogenic heating. 

The melt fraction is the ratio of the planet's mass in molten form to the planet's total mass. 
\citet{abe1997thermal} parameterized the melt fraction as
\begin{align} 
    \phi = \dfrac{T_{\text{p}} - T_{\text{solidus}}}{T_{\text{liquidus}} - T_{\text{solidus}}}
    \label{eqn:melt_fraction}
\end{align}
where $T_{\text{solidus}}$ is the solidus temperature and $T_{\text{liquidus}}$ is the liquidus temperature of the mantle given by $T_{\text{\text{liquidus}}} = T_{\text{\text{solidus}}} + 600$ K \citep{andrault2011solidus}.
While the mantle potential temperature varies with time, $T_{\text{liquidus}}$ and $T_{\text{solidus}}$ vary with radius.
In order to show the variation of melt fraction for the mantle, we chose $T_{\text{\text{solidus}}} = 2076.26$ K which denotes the mean solidus for the upper mantle.
In this study, we do not model the evolution of the solidus or liquidus. Traditionally, the melt fraction has been calculated at the surface \citep{barth2021magma, krissansen2021oxygen}. However, we opted to calculate the melt fraction at the mean solidus due to the substantial temperature difference ($\sim 700$ K) between the surface and the mean solidus.
As the melt fraction reaches a critical value ($\phi_{\text{c}}$=0.4), the viscosity changes drastically \citep{abe1997thermal, lebrun2013thermal}.
The viscosity ($\eta$) of the magma ocean can be parameterized as a function of $T_{\text{p}}$ and $\phi$ as follows \citep{lebrun2013thermal}
\begin{align} 
    \eta=\begin{cases}
        { \dfrac{2.4 \times 10^{-4} \, \exp \left( \dfrac{4600}{T_{\text{p}} - 1000} \right)}{ \left( \dfrac{\phi - \phi_{c}}{ 1 - \phi_{c}} \right)^{2.5}}} \qquad \qquad & {\phi>\phi_{c}} \\ \\
        { 3.8 \times 10^{9} \,\exp \left( \dfrac{3.5 \times 10^5}{RT_{\text{p}}} \right)} \qquad \qquad & {\phi \leq \phi_{c}}\\
    \end{cases}
    \label{eqn:viscosity}
\end{align}
Below $\phi_{\text{c}}$, the viscosity behaves like a solid, leading the planet to undergo solid-state evolution.

\subsection{Modeling the atmosphere} \label{subsec:atmosphere}

\texttt{SERPINT} utilizes a grey atmosphere framework inspired by \citet{elkins2008linked}, a simple approach for computing outgoing long-wave radiation. 
The grey radiative transfer method aggregates the average infrared opacities of each gas species, adjusted according to their respective partial pressures, rather than computing opacity for each wavelength bin individually. 
The atmospheric net flux is determined by the Stefan-Boltzmann emission from the thin atmosphere of the planet, which can be written as \citep{elkins2008linked}
\begin{align}
    F=\lambda \sigma \left(T_{\text{p}}^{4}-T_{\text{eq}}^{4}\right)
    \label{eqn:flux2}
\end{align}
Here $T_{\text{p}}$  is the mantle potential temperature and $T_{\text{eq}}$ is the blackbody equilibrium temperature of the planet heated only by the star, $\sigma$ is the Stefan-Boltzmann constant, and $\lambda$ is the emissivity. 
We have chosen $\lambda=0.2$ for the model \citep{barth2021magma}. 
The planet equilibrium temperature can be calculated as follows
\begin{align}
    T_{\text{eq}}&=\left(\dfrac{\mathcal{F}_\star\left(1-A_{\text{B}}\right)}{4\sigma}\right)^{1/4}
    \label{eqn:T_eq}
\end{align}
where $\mathcal{F}_\star$ is the stellar flux and $A_{\text{B}}$ is the bond albedo. $\mathcal{F}_\star$ can be written according to Eqn. \ref{eqn:flux_star} \citep{miller2011heavy}
\begin{align} 
    \mathcal{F}_\star&=\dfrac{L_\star}{4\pi a^{2} \sqrt{1-e^2}}
    \label{eqn:flux_star}
\end{align}
where $L_\star$ is the luminosity of the star, $a$ is the semi-major axis of the orbit, and $e$ is the eccentricity of the orbit.

\subsection{Solid state evolution} \label{subsec:solid_state_evol}

After the melt fraction drops below $\phi_{\text{c}}$, the planet enters solid-state evolution regime.
In this regime, the mantle, despite being solid, behaves like a highly viscous fluid and convects over geologic timescales ($\sim$few Myrs) with the adiabatic mantle potential temperature, $T_{\text{p}}$.
The outgoing flux decreases from the Stefan-Boltzmann emission flux to solid-state flux \citep{lebrun2013thermal}, described as
\begin{equation}
    q_s = C \dfrac{k (T_{\text{p}} - T_{\text{surf}})}{d} \left( \dfrac{Ra}{Ra_{\text{crit}}} \right)^{1/3}
    \label{eqn:q_s}
\end{equation}
Here $T_{\text{surf}}$ is the surface temperature in equilibrium with the atmosphere, $k$ is the thermal conductivity of the mantle (4.2 W m$^{-1}$ K$^{-1}$), $d$ is the mantle thickness of the planet, $Ra$ is the Rayleigh number and $C$ (2.12035, dimensionless) is a constant. 
To evaluate this constant, we considered the case of Earth, in our model, with a net heat outflux of 0.1 W m$^{-2}$ from the Earth's surface \citep{davies2010earth} when the mantle potential temperature reaches a reference temperature of 1600 K \citep{foley2020heat}. The value obtained is the constant which we use to calculate the heat outflux for this planet.
The Rayleigh number ($Ra$) is a dimensionless parameter that represents the ratio of forces driving convection, primarily from thermal buoyancy, to the forces resisting it, such as thermal diffusivity and viscosity.
It is defined as 
\begin{equation}
    Ra = \dfrac{\rho_{\text{m}} g_{\text{s}} d^3 \alpha (T_{\text{p}} - T_{\text{surf}})}{\kappa \eta}
\end{equation}
where $\rho_{\text{m}}$ is the mantle density, $\alpha$ is the thermal expansion coefficient (2 $\times$ 10$^{-5}$ K$^{-1}$), $g_{\text{s}}$ is surface gravity, $\kappa$ is the thermal diffusivity of the mantle (10$^{-6}$ m$^2$ s$^{-1}$) and $\eta$ is the viscosity obtained from Eqn. \ref{eqn:viscosity}.
For a fluid to undergo convection, it must have a Rayleigh number higher than 1000, which is also called as the critical Rayleigh number ($Ra_{\text{crit}}$) \citep{trompert1998rayleigh, tasker2020planetary}.
Finally, this heat flux, $q_s$, can be substituted for $F$ in Eqn. \ref{eqn:dT_p_dt} for the case $\phi < \phi_{\text{c}}$.

\subsubsection{Volatile model} \label{subsubsec:volatile}

The volatile model regulates water and oxygen exchange within the planet's reservoirs, i.e., the liquid melt, the solid part, and the atmosphere.
We begin the simulation with 10 earth oceans (EO) of water dissolved in the melt.
Initially, it is assumed that all the water dissolves into the magma ocean. 
The water mass balance for the system is given by \citep{schaefer2016predictions}.
\begin{align}
    M_{\text{H$_2$O}}^{\text{magma}} = M_{\text{H$_2$O}}^{\text{solid}} + M_{\text{H$_2$O}}^{\text{liquid}} + M_{\text{H$_2$O}}^{\text{atm}}
    \label{eqn:M_H2O_magma_1}
\end{align}
More specifically, we can rewrite this equation as 
\begin{gather}
\begin{aligned}
    M_{\text{H$_2$O}}^{\text{magma}} = & k_{\text{H$_2$O}}f_{\text{H$_2$O}}M^{\text{solid}} + f_{\text{H$_2$O}}M^{\text{liquid}} \\
    & + \dfrac{4\pi R_{\text{p}}^2}{g}P_{\text{H$_2$O}}
    \label{eqn:M_H2O_magma_2}
\end{aligned}
\end{gather}
where $M_{\text{H$_2$O}}^{\text{solid}}$, $M_{\text{H$_2$O}}^{\text{liquid}}$ and $M_{\text{H$_2$O}}^{\text{atm}}$ represent the water mass present in the solidified mantle, liquid melt and atmosphere, respectively.
$f_{\text{H$_2$O}}$ and $k_{\text{H$_2$O}}$ denote the water mass fraction and mantle-averaged melt-solid water partition coefficient, respectively.
We have used $k_{\text{H$_2$O}} = 0.01$, similar to \citet{schaefer2016predictions}.
As the magma ocean cools, solidification begins from the bottom, which traps water in solid rock, retaining only a fraction of the dissolved water. 
At equilibrium, we model the water mass fraction in the melt as a function of its partial pressure. 
We use the \citet{duan2014general} fit to establish the relationship between the water fraction ($f_{H_2O}$) and partial pressure of water ($P_{H_2O}$) at higher pressures than commonly used results from \citet{papale1999modeling}. 
The fitted equation can be written as follows 
\begin{align}
    P_{\text{H$_2$O}} (	\text{GPa}) = 31.855 (f_{\text{H$_2$O}} + 0.0523)^{2.15} - 0.056
    \label{P_H2O}
\end{align}

During solidification, most water moves to the melt, as solid rocks cannot hold large amounts of water, increasing the water mass fraction of the magma ocean, as depicted by Eqn. \ref{eqn:dM_H2O_sol} and Eqn. \ref{eqn:dM_H2O_magma} \citep{schaefer2016predictions}.

\begin{align}
    \dfrac{dM_{\text{H$_2$O}}^{\text{solid}}}{dt} = 4\pi \rho_{\text{m}} k_{\text{H$_2$O}}f_{\text{H$_2$O}}r_{\text{s}}^2\dfrac{dr_{\text{s}}}{dt}
    \label{eqn:dM_H2O_sol}
\end{align}
\begin{align}
    \dfrac{dM_{\text{H$_2$O}}^{\text{magma}}}{dt} = -\dfrac{dM_{\text{H$_2$O}}^{\text{solid}}}{dt} - 4\pi R_{\text{p}}^2 \:\Phi_1\: \dfrac{\mu_{\text{H$_2$O}}}{2 \mu_\text{H}} 
    \label{eqn:dM_H2O_magma}
\end{align}
where $\mu_{\text{H}}$ and $\mu_{\text{O}}$ are the atomic masses of hydrogen and oxygen, respectively, and $\Phi_1$ is the escape flux rate of hydrogen.

Water is outgassed into the atmosphere to restore the pressure equilibrium, raising atmospheric pressure.
Due to stellar XUV radiation, water gets photolyzed into hydrogen and oxygen, where all hydrogen and some oxygen escape into space. 
The escape flux rate of oxygen is denoted by $\Phi_2$.
The remaining oxygen present in the atmosphere starts dissolving into the magma ocean, leading to the oxidation of FeO to Fe$_2$O$_3$. Once FeO is fully oxidized, oxygen accumulates in the atmosphere.
An analogous set of equations regulates the quantity of oxygen across the reservoirs \citep{schaefer2016predictions}:
\begin{align}
    \dfrac{dM_{\text{O}}^{\text{solid}}}{dt} = 4\pi \rho_{\text{m}} f_{\text{FeO$_{1.5}$}}r_{\text{s}}^2\dfrac{dr_{\text{s}}}{dt} \dfrac{\mu_\text{O}}{2\mu_{\text{FeO$_{1.5}$}}}
    \label{eqn:dM_O_sol}
\end{align}
\begin{align}
    \dfrac{dM_{\text{O}}^{\text{magma}}}{dt} = -\dfrac{dM_{\text{O}}^{\text{solid}}}{dt} + 4\pi R_{\text{p}}^2 \left( \Phi_1 \dfrac{\mu_\text{O}}{2\mu_\text{H}} - \Phi_2 \right)
    \label{eqn:dM_O_magma}
\end{align}

We have adopted a zero-dimensional model for our atmosphere, which assumes that all atmospheric constituents mix uniformly and instantaneously throughout the entire atmosphere. 
Consequently, the dissociated molecules in the upper part of the atmosphere due to XUV radiation can reach the bottom of the atmosphere freely and get dissolved back into the melt, oxidizing the mantle.
Similarly, water released from the surface can access all the XUV radiation \textit{}at the top of the atmosphere. This allows for its immediate photolysis without interference from the produced hydrogen and oxygen.

To keep our model simple, we assume that as long as the magma ocean's FeO reservoir is not completely oxidized, no oxygen will accumulate in the atmosphere. 
If the mantle initially contains a low amount of volatiles and the escape is too strong, the planet eventually loses all its volatiles in a matter of time.

\subsubsection{Escape parameterizations} \label{subsubsec:escape}

The primary mechanism for atmospheric loss is the action of XUV radiation and stellar activity, which washes away lighter molecules, such as hydrogen and helium, from the atmosphere.
XUV radiation from M-dwarfs is higher in the first million years, which reduces as the star evolves and cools down.
The XUV luminosity is calculated using Eqn. \ref{eqn:XUVlum}.
For M-dwarfs, it is observed to be 1000 times lower than their bolometric luminosity \citep{luger2015extreme}.
Moreover, due to their small size, stellar activities on M-dwarfs show more significant luminosity variations, and often last longer than higher-mass stars.
In such cases, vigorous stellar winds and outbursts can erode the atmospheres significantly.
However, our model does not have stellar activities due to their stochastic and chaotic nature.
The XUV-driven hydrodynamic escape flux is given as \citep{zahnle1990mass, luger2015extreme}
\begin{align}
    \Phi = \dfrac{\epsilon \mathcal{F}_{\text{XUV}} R_{\text{p}}}{4GM_{p}} \qquad \text{and, } \qquad \Phi^{\text{ref}}_{\text{H}} = \dfrac{\Phi}{m_{\text{H}}}     
    \label{eqn:ref_flux_H}
\end{align}
where $\mathcal{F}_{\text{XUV}}$ is the stellar XUV flux, $\epsilon$ is the XUV absorption efficiency, $R_{\text{p}}$ is the planet radius, $M_{\text{p}}$ is the mass of the planet, $m_i$ is the mass of $i^{th}$ atom species, and $\Phi^{\text{ref}}_{\text{H}}$ is the reference hydrogen escape flux calculated when all the escaping mass is considered only as hydrogen. We have taken $\epsilon=0.3$ for our calculations \citep{luger2015extreme}.
For individual species, one can write
\begin{gather}
    m_{\text{H}} \Phi^{\text{ref}}_{\text{H}} = \sum_i m_i \Phi_i
    \label{eqn:flux_H_individual}
\end{gather}
Considering the escape of only H and O species, we obtain the relation
\begin{gather}
    m_{\text{H}} \Phi_{\text{H}}^{\text{ref}} = m_{\text{H}} \Phi_{\text{H}} + m_{\text{O}} \Phi_{\text{O}}
    \label{eqn:flux_H_O}
\end{gather}
where
\begin{align}
    \Phi_{\text{O}} = \dfrac{X_{\text{O}}}{X_{\text{H}}} \Phi_{\text{H}} \left( \dfrac{m_{\text{c}} - m_{\text{O}}}{m_{\text{c}} - m_{\text{H}}} \right)
    \label{eqn:flux_O}
\end{align}
where $X_{\text{O}}$ and  $X_{\text{H}}$  is the molar mixing ratio of oxygen and hydrogen, 
respectively \citep{hunten1973escape, chassefiere1996hydrodynamic}.

\citet{hunten1987mass} demonstrated in their study on mass fraction associated with hydrodynamic escape that an escaping species can effectively drag along a heavier species. This occurs if the latter's mass is less than the crossover mass, $m_{\text{c}}$, under the influence of the atomic hydrogen flow. The crossover mass can be written as
\begin{align}
    m_{\text{c}} = m_{\text{H}} + \dfrac{k_B T}{b_{\text{HO}} g_{\text{s}} X_{\text{H}}} \Phi_{\text{H}}
    \label{eqn:crossover_mass_1}
\end{align}
here, $b_{\text{HO}}$ is the binary diffusion coefficient for the two species H and O, $g_{\text{s}}$ is the surface gravity \citep{hunten1973escape, chassefiere1996hydrodynamic}. 
We utilized $b_{\text{HO}} = 4.8 \times 10^{17}$ (T/K)$^{0.75}$ cm$^{-1}$ s$^{-1}$ \citep{zahnle1986photochemistry}. 
The value $T = 400$ K denotes the average thermospheric temperature of the outflowing species \citep{hunten1987mass, chassefiere1996hydrodynamic}.
The total energy-limited mass loss rate, i.e., the combined upward escape rates of hydrogen and oxygen, is defined as
\begin{gather}
    \dot{M}_{EL} = 4 \pi R_{\text{p}}^2 \Phi_{\text{H}}^{\text{ref}} m_{\text{H}}
    \label{eqn:M_dot_EL}
\end{gather}
Here, we assume that the net oxygen and hydrogen result from the photolysis of the atmospheric water. 
It also assumes that the dissociation of H$_2$ and O$_2$ occurs rapidly enough for both species to exist predominantly in atomic form near the lower region of the flow \citep{chassefiere1996hydrodynamic}.
More information on the escaping mass relations can be found in appendix \ref{apdx:sec:escape}.

The critical XUV flux is defined as the threshold below which hydrodynamic escape of oxygen is not possible.
It is calculated using the crossover mass $m_{\text{c}}$ by the following equation
\begin{align}
    \mathcal{F}_{\text{XUV}}^{\text{crit}} = \dfrac{4G M_{\text{p}} b_{\text{HO}} g_{\text{s}}}{k_B T \epsilon R_{\text{p}}} X_{\text{H}} (m_{\text{c}} - m_{\text{H}}) m_{\text{H}}
    \label{eqn:crossover_mass_2}
\end{align}
Species with molecular masses less than the crossover mass are subject to hydrodynamic escape from the atmosphere.
For the hydrodynamic escape of oxygen, we get the crossover mass as  $m_{\text{c}} \geq m_{\text{O}}$ \citep{luger2015extreme}.
\begin{align}
    m_{\text{c}} = \dfrac{43}{3} m_{\text{H}} + \dfrac{1}{6} \dfrac{k_B T \Phi_{\text{H}}^{\text{ref}}}{b_{\text{HO}} g_{\text{s}}}
    \label{eqn:XUV_flux_crit}
\end{align}

By setting $m_{\text{c}} = m_{\text{O}}$ and solving Eqn. \ref{eqn:crossover_mass_2} and \ref{eqn:XUV_flux_crit}, we get the lower bound for the reference flux to cause the hydrodynamic escape of oxygen as
\begin{gather}
    \Phi_{\text{H}}^{\text{ref}} \geq \dfrac{10 m_{\text{H}} b_{\text{HO}} g_{\text{s}}}{k_B T}
    \label{eqn:ref_flux_H_2}
\end{gather}
Here, we have employed $m_{\text{O}}$ = 16 $m_{\text{H}}$ to simplify things. 
We also used $X_{\text{H}} = 2/3$ and $X_{\text{O}} = 1/3$, which relates to the typical molar mixing ratio for a steam atmosphere, under the assumption that all hydrogen and oxygen are produced via photolysis.

\section{Results} \label{sec:results}

This section presents our findings on the interior structure, as well as the thermal and atmospheric evolution of GJ 486b, using the 1-D self-consistent coupled model \texttt{SERPINT}. By calculating the planet's core-radius fraction, density profile, and thermal evolution, we provide insights into the early stages of GJ 486b's interior and its subsequent cooling and solidification. Our results offer a comprehensive understanding of the complex interactions governing the evolution of GJ 486b's interior and atmosphere.

\begin{figure*}[ht]
\centering
\includegraphics[width=\textwidth]{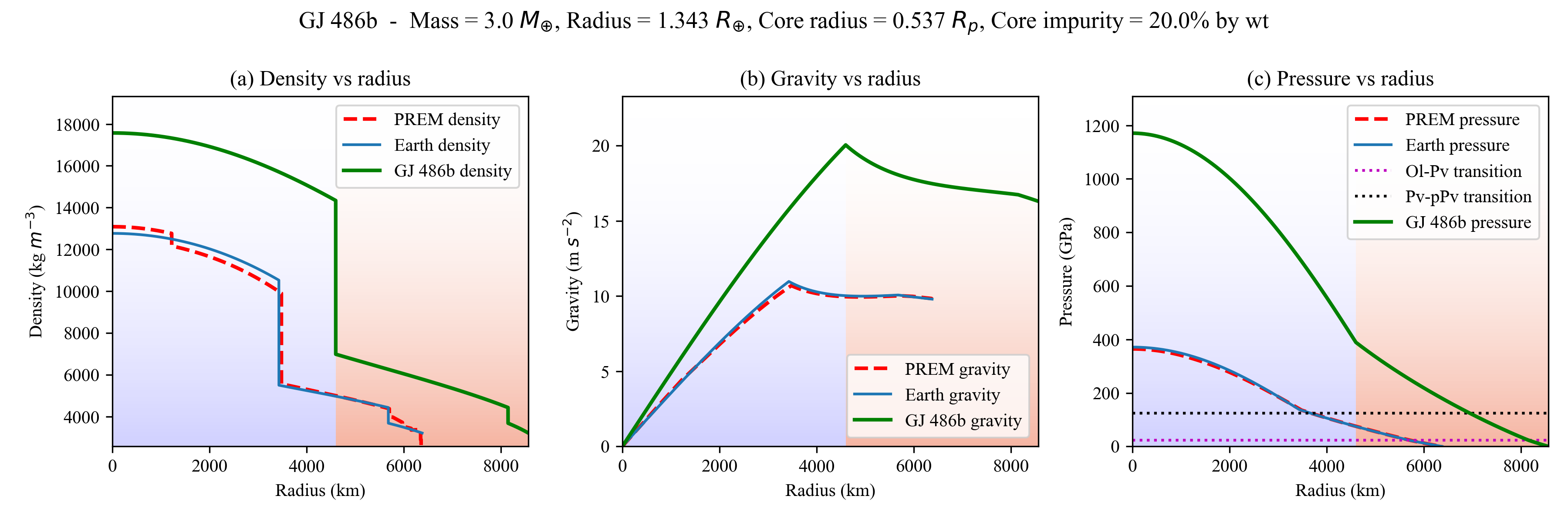}
\caption{Comparison of the interior structure of GJ 486b \textit{(green)} with the PREM \textit{(red dashed)} and Earth \textit{(blue)}. The plots show variations in density, gravity, and pressure, arranged from left to right. The bluish and brownish background colors denote the core and the mantle of GJ 486b, respectively.}
\label{fig:Earth_and_GJ486b_structure}
\end{figure*}

\subsection{Interior structure of Earth} \label{subsec:result_structure_earth}

Our structural model assumes a simplified interior composition similar to that of Earth's. The hydrostatic equilibrium equations are used to determine the planet's structure, incorporating the EOS for different layers of the planet's interior. We initialize the model with a preexisting iron core containing impurities, allowing the model to calculate the core radius using the EOS and then solve for the mantle structure.

\begin{figure*}[ht]
\centering
\includegraphics[width=\textwidth]{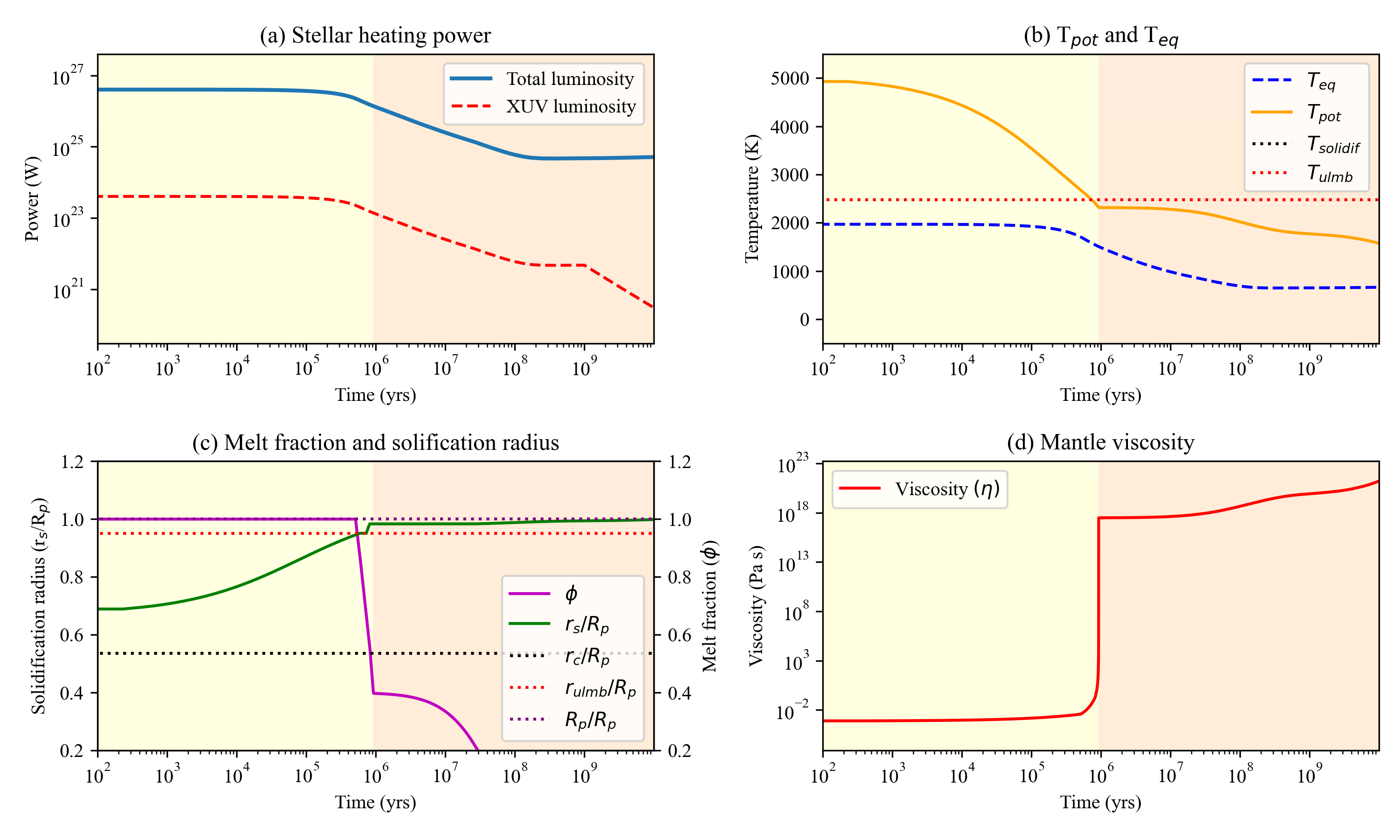}
\caption{Evolution of thermal properties of GJ 486b. The two plots at the top denote the stellar luminosity evolution (bolometric and XUV) and the mantle potential temperature evolution for GJ 486b. The yellow and orange background denote the time evolution before and after the complete solidification of the planet, respectively. The bottom two plots illustrate the planet radius, melt fraction, and viscosity evolution till the point of complete solidification. The dotted lines denote the core radius ($r_c$), upper-lower mantle boundary radius ($r_{\text{ulmb}}$) and the planet radius ($R_p$), all normalized to planet radius.}
\label{fig:GJ486b_thermal}
\end{figure*}

Figure \ref{fig:Earth_and_GJ486b_structure} presents our model's results for Earth, validated against PREM \citep{dziewonski1981preliminary}. PREM offers detailed information on Earth's composition, density, and physical properties, serving as a fundamental reference for understanding Earth's internal dynamics.

In our model, we assume a 20\% by weight core impurity, comprising 12\% silicon, 7\% oxygen, and 1\% sulfur. This assumption yields results that align well with PREM data for density, gravity, and pressure, with only slight differences observed. Specifically, we determine Earth's core-radius fraction to be approximately 0.538 $R_{\oplus}$ (3429 km), whereas PREM suggests about 0.546 $R_{\oplus}$ (3485 km).

The density values from our model are as follows: 12.77 kg m$^{-3}$ at Earth's center, 5.50 kg m$^{-3}$ near the core-mantle boundary, and 3.69 kg m$^{-3}$ near the upper-lower mantle boundary. These numbers show minimal deviation from PREM's density profile. The discrepancies can be attributed to our model's assumption of a completely liquid core, in contrast to Earth's actual structure, which includes a solid inner core and a liquid outer core. This distinction leads to a density discontinuity at approximately 1220 km, as depicted in Fig. \ref{fig:Earth_and_GJ486b_structure}(a), and accounts for the slight differences in core radius and density profile compared to PREM.

Furthermore, minor variations in the upper mantle's profile can be attributed to the complex structure of Earth's crust and upper mantle, which consist of diverse materials and minerals formed over billions of years of tectonic activity. These complexities are not fully captured in our current model, leading to the observed differences.

Based on the density profile, we calculate the gravity profile using Eqn. \ref{eqn:grav} and observe that the interior gravity increases initially in the core, reaching up to 10.97 m s$^{-2}$, and then decreases in the mantle, reaching a value of 9.80 m s$^{-2}$ at the surface, as shown in Fig. \ref{fig:Earth_and_GJ486b_structure}(b). The main reason for this is the significant density difference between the core and the mantle. Similarly, the pressure increases gradually through the mantle, reaching 138 GPa at the core-mantle boundary, and then rises steeply within the core, reaching a pressure of 371 GPa at Earth's center, as shown in Fig. \ref{fig:Earth_and_GJ486b_structure}(c). Both the gravity and pressure profiles closely match those from PREM, validating our model.


\subsection{Interior structure of GJ 486b} \label{subsubsec:result_structure_GJ486b}


With an interior structure similar to Earth, our model predicts that the rocky planet GJ 486b has a core 1.34 times larger than Earth's core. It presents a core-radius fraction of 0.537 ($R_{\text{p}}$), corresponding to 4600 km with a 20\% core impurity. 
\citet{meier2024geodynamics} use the planetary parameters of GJ 486b as 2.81 $M_{\oplus}$ and 1.31 $R_{\oplus}$ with an Earth-like interior composition \citep{trifonov2021nearby}, leading to a core size of 0.586 $R_{\oplus}$, corresponding to 4900 km. The deviation in core size is due to the difference in the observed mass and radius of the planet.
The radial variations of density, gravity, and pressure structures of GJ 486b compared to the Preliminary Reference Earth Model (PREM) can be seen in Fig. \ref{fig:Earth_and_GJ486b_structure}(a). 
Due to the three times higher mass of GJ 486b, the density at the center of the planet reaches 17.57 kg m$^{-3}$, while the density at the core-mantle boundary and the upper-lower mantle boundary reaches 6.98 kg m$^{-3}$ and 3.69 kg m$^{-3}$, respectively, as shown in Fig. \ref{fig:Earth_and_GJ486b_structure}(b). The interior gravity of GJ 486b initially increases in the core to 20.05 m s$^{-2}$ and then decreases to reach a final value of 16.30 m s$^{-2}$ at the planet's surface. Similarly, the pressure initially increases to 23.5 GPa at the upper-lower mantle boundary, denoting the olivine-perovskite (Ol-Pv) transition, then to 125 GPa denoting the perovskite-post-perovskite (Pv-pPv) transition, and then increases to 390 GPa at the core-mantle boundary, finally reaching a substantial value of 1171 GPa at the center of the planet, as shown in Fig. \ref{fig:Earth_and_GJ486b_structure}(c).

\subsection{Thermal and atmosphere evolution of GJ 486b} \label{subsec:result_thermal}


The model initializes with the beginning of the evolution of both the star and the planet. GJ 486, an M3V-type star with a mass of 0.323 $M_{\odot}$, begins with a peak bolometric luminosity of $4.02 \times 10^{26}$ W m$^{-2}$ and remains stable for about 0.2–0.3 Myrs before decreasing. The stellar evolution is governed by the stellar models from \citet{baraffe2015new}. Fig. \ref{fig:GJ486b_thermal}(a) illustrates the evolution of bolometric and XUV luminosity for GJ 486. The XUV luminosity is 1000 times less than the bolometric luminosity, governed by $\beta_{XUV}$. After one billion years, when it exceeds the saturation time, the XUV luminosity decreases according to Eqn. \ref{eqn:XUVlum}.

We initialize \texttt{SERPINT} with a mantle potential temperature of 5000 K, ensuring it is entirely molten. 
Subsequently, the model cools down towards the mantle’s equilibrium temperature. However, the extent of the magma ocean depends on the solidus of the planet, which also depends on its size.
Concurrently, heat sources such as radioactive decay and stellar irradiation extend the duration of the magma ocean.
Fig. \ref{fig:GJ486b_thermal}(b) illustrates how the mantle potential temperature (orange) evolves with time towards the equilibrium temperature (blue dashed) of the planet. 
In this case, the simulation initially begins with a 14\% solid mantle owing to the high solidus temperature of the core-mantle boundary region.
The mantle then solidifies from the bottom as the mantle potential temperature decreases with time, following the mantle solidus profile. 
The cooling rate decelerates as solidification proceeds, primarily because the latent heat released during crystallization slows the cooling process.
The solidification radius evolves from $r_s=0.68$ $R_{\text{p}}$ at $T_{\text{p}}=5000$ K to $r_s=0.95$ $R_{\text{p}}$ at the $T_{\text{p}}=2600$ K where it touches the upper-lower mantle boundary in a span of 0.59 Myrs.
At this point, the lower mantle gets completely solidified, whereas the solidification of the upper mantle commences significantly later. 
This discrepancy arises from the two layers' differing mineral compositions and solidification temperatures, as described in \ref{subsec:composition}.
The solidification radius stays at the upper-lower mantle boundary till the temperature cools enough to reach the solidus temperature of the upper mantle of 2475 K.
Fig. \ref{fig:GJ486b_thermal}(c) (green) shows how the solidification radius evolves and advances further in the upper mantle.

Simultaneously, the melt fraction (Fig. \ref{fig:GJ486b_thermal}(c), magenta) of the planet starts to decrease, and the mantle viscosity begins to rise (Fig. \ref{fig:GJ486b_thermal}(d)). The melt fraction and viscosity illustrate how the rheology of the planet changes. As the planet cools, the melt slowly forms crystals of minerals that grow in size over time. This causes thickening of the melt, gradually increasing its viscosity and forming a mush layer, which solidifies to form rocks. As the melt fraction reaches the critical value of 0.4, the viscosity increases drastically from a low-viscosity fluid to a high-viscosity solid-like state after about 0.93 Myrs. At this point, we consider the mantle to be primarily solidified, denoted by the orange shades in the plots of Fig. \ref{fig:GJ486b_thermal}. As a result, the planet undergoes solid-state convection, and the heat outflux drops by more than 1000 times. This slows down the cooling rate, and the mantle potential temperature takes longer to decrease.
This can also be seen in the form of a sudden increase in the radius of solidification in the upper mantle, as shown in Fig. \ref{fig:GJ486b_thermal}(c). As the rate of solidification of the planet slows down, the rate of decrease of the melt fraction also decreases. The planet continues to cool down thereafter towards its equilibrium temperature, reaching a minimum of 1574 K by the end of the simulation (10 Gyrs).

\subsection{Volatile and escape evolution of GJ 486b} \label{subsec:result_atm}

\begin{figure*}[ht]
\centering
\includegraphics[width=\textwidth]{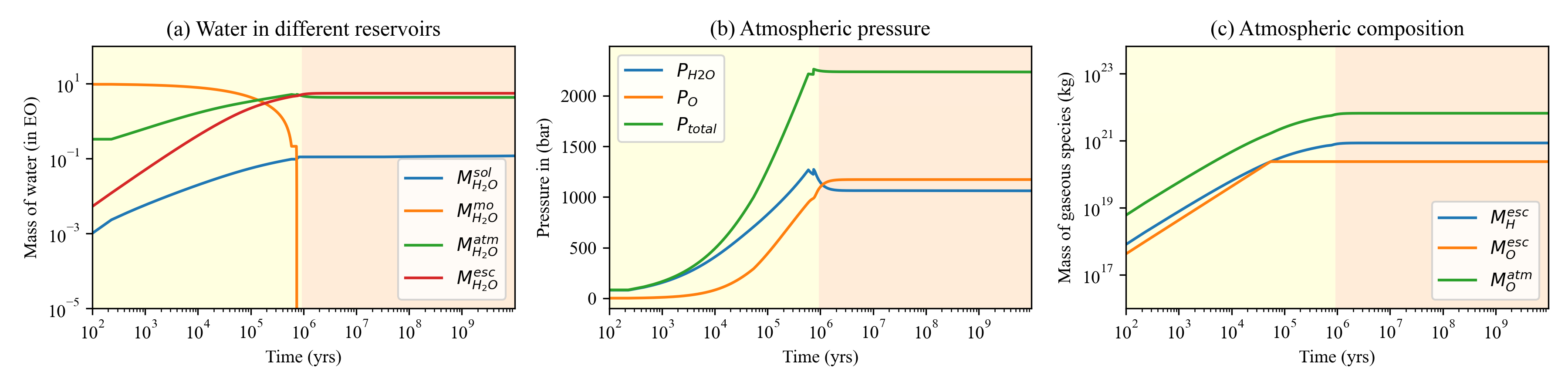}
\caption{Evolution of volatiles in different reservoirs of GJ 486b. The plot on the left illustrates the water distribution among different reservoirs. The middle plot shows the pressure buildup of water and oxygen in the atmosphere. The plot on the right shows the mass of escaping volatiles from the atmosphere. The yellow and orange background denotes the time evolution before and after the complete solidification of the planet, respectively.}
\label{fig:GJ486b_atm}
\end{figure*}


Initially, most of the volatiles on the planet, particularly water in this instance, are assumed to be dissolved in the melt. We started the simulation with a total of 10 Earth oceans (EO) of water. A small amount of water remains in the atmosphere due to the melt-vapor equilibrium, which is controlled by the pressure from the melt and solubility relations, as shown in Eqn. \ref{P_H2O}. During solidification, water is transferred from the melt to the solid phase and the atmosphere, resulting in an increase in the amount of water in the solid and in the atmospheric water pressure. This transfer causes a decrease in the amount of water in the melt, primarily due to the reduction in the amount of melt as the planet cools and solidifies.

Concurrently, atmospheric water dissociates into hydrogen and oxygen. Since we assume a simplified model for the atmosphere, we consider a uniform mixing of dissociated molecules and atoms. This means that the oxygen atoms dissociated at the top of the atmosphere can reach the bottom, where they get ingested into the melt and oxidize the mantle FeO to Fe$_2$O$_3$. Upon complete solidification, the outgassing of water ceases, and only atmospheric escape drives the movement of water within the planet’s reservoirs.


Fig. \ref{fig:GJ486b_atm}(a) illustrates the distribution of water across different reservoirs considered in the model. We observe that the total water summed across all reservoirs equals the total water present at the beginning of the simulation. 
As solidification proceeds, the amount of water in the atmosphere increases, causing the atmospheric pressure of water to rise from 61.5 bars to 1275.8 bars around $10^5-10^6$ years. 
As the lower mantle solidifies and the radius of solidification halts, the water pressure begins to drop. However, it increases again as the planet starts to solidify rapidly, forming an M-shape in Fig. \ref{fig:GJ486b_atm}(b).
Similar set of results have been shown by \citet{barth2021magma} and \citet{bower2022retention}, where the atmospheric water pressure increased from a few bars to a few hundred bars in about 1 Myrs for Earth-sized planets.
\citet{bower2022retention} investigated the atmospheres of Earth-like magma ocean planets for a range of oxygen fugacities, C/H ratios, and hydrogen budgets. 
They show that for equilibrium crystallization of magma oceans, more water is retained within the magma ocean leading to a carbon-rich atmosphere. 

Fig. \ref{fig:GJ486b_atm}(b) also illustrates the pressure buildup of oxygen in the atmosphere. As time passes and more water gets photolyzed, the water pressure drops and eventually reaches a nearly stable condition around the time of complete solidification of the planet, when the XUV radiation is no longer strong enough to cause further dissociation of water. The XUV radiation is also insufficient to facilitate the escape of oxygen, leading to an accumulation of oxygen in the atmosphere. The escaping masses of hydrogen and oxygen, along with the accumulated oxygen mass in the atmosphere, are represented in Fig. \ref{fig:GJ486b_atm}(c).

\section{Discussion} \label{sec:discussion}

\subsection{New Insights from \texttt{SERPINT} for GJ 486b in the context of recent JWST Observations} \label{subsec:analysis}

We simulated the interior structure of GJ 486b along with its thermal and atmospheric evolution using \texttt{SERPINT}.
Our results predict a core radius 1.34 times larger than that of the Earth. A higher planet mass and core size implies a higher interior pressure. 
We observed the planet's central pressures as high as 1171 GPa, which is more than three times the Earth's central pressure, i.e., about 371 GPa.
The EOSs used in \texttt{SERPINT} are based on the presence of $\varepsilon$-Fe at such extreme pressures. 
Detailed studies of materials at such high pressures in laboratories is challenging with the present technology. 
\citet{smith2018equation} investigated the behavior of solid iron at pressures reaching up to 1.4 TPa, maintaining a fixed temperature of 965 K. \citet{kraus2022measuring} studied the behavior of both solid and liquid iron across a wide range of temperatures, extending up to 1 TPa. 
These studies have significantly contributed to our understanding of material behavior at high pressures.
Nonetheless, the validations of \texttt{SERPINT} with PREM provides safeguards for the assumptions to hold true.
Slight differences occur in the core-radius fraction and density profile for Earth compared to PREM \citep{dziewonski1981preliminary}, primarily due to our assumption of a fully liquid iron core; whereas PREM accounts for Earth's solid inner core and liquid outer core.
Despite these variations, the gravity and pressure profiles generated by our model are consistent with PREM to very good extent, validating our overall approach.

The thermal evolution model for GJ 486b illustrates the planet’s cooling and solidification from an initially molten state.
The planet is initialized with a mantle potential temperature of 5000 K and part of the lower mantle already solidified. 
This is because the core-mantle boundary of GJ 486b is present at a depth of 3966.8 km from the surface, corresponding to a temperature of 11747 K for the core-mantle boundary region to get molten. 
A temperature of 5000 K starts the solidification at a depth of 2663.9 km which is 0.689 $R_{\text{p}}$.
The Earth's core-mantle boundary on the other hand, which is at a depth of 2948.6 km starts around the same depth, with an initially solidified radius of 3569.2 km which is 0.559 $R_{\oplus}$. 

The hot mantle slowly cools and the solidification radius grows towards the surface.
The melt fraction decreases correspondingly, reflecting the planet's transition from a predominantly molten state to a solidified mantle.
The planet undergoes a jump in its viscosity over a span of approximately 0.93 million years which denotes the transformation from liquid state to solid state convection, i.e., its solidification. 
The upper mantle solidifies slowly due to its steeper solidus profile, as well as reduced cooling rate due to solid-state convection, which affects the overall thermal evolution. This discrepancy underscores the impact of compositional differences and solidus temperatures on the thermal evolution of the planet. 
In the context of solid-state convection, the heat outflux experiences a significant reduction, thereby extending the cooling duration. The Earth's outflux of 0.1 W m$^{-2}$ \citep{davies2010earth} at 1600 K is utilized as a reference to determine the constant (C) in Eqn \ref{eqn:q_s}. 
Lower heat fluxes, such as 0.02 W m$^{-2}$ observed in planets like Mars \citep{parro2017present}, result in a fivefold decrease in C and a substantial reduction in cooling rates. 
Conversely, increasing the heat flux by a factor of five leads to a notable acceleration in cooling. 
However, it is important to note that these effects are also highly dependent on additional factors, including planetary mass, size, and composition.
The large size of the planet is also expected to reduce the increment in solidus temperature with pressure in the lower mantle \citep{fei2021melting}.
However, such effects are beyond the scope of this version of \texttt{SERPINT}.
We have used a no-flux boundary condition at the core-mantle boundary, which is a common and practical approach when focusing on the internal processes and dynamics without external influences. It simplifies the model by assuming that there is no exchange of material or heat across the core-mantle boundary, which is appropriate for this case, since we do not account for core-heating in this model.

Being higher in mass, larger in size, and closer to its host star, it is possible that GJ 486b drives mantle convection patterns and core dynamics very different from that on Earth, which may lead to a different solidifying timescale compared to Earth. 
\citet{tackley2013mantle} observed that super-Earths may exhibit super-adiabatic and isoviscous deep mantles.
Studies by \citet{van2011mantle, meier2021hemispheric} have demonstrated the possibility of hemispheric tectonics on hot exoplanets, such as LHS 3844b, due to differing convection properties between the dayside and nightside of the planet. A strong lithosphere is likely to result in downwelling on one hemisphere and upwelling on the other \citep{meier2021hemispheric, meier2024geodynamics}.
Larger planets have more volumes compared to surface areas, which enables them to hold more heat compared to smaller planets \citep{foley2020heat}.
This suggests that they should take more time to solidify compared to the Earth. 
\citet{hamano2013emergence} showed this effect on the solidification of the planet due to its mass and orbital distance.
However, due to higher mantle potential temperatures, they might also have more volcanic and tectonic activities causing high rates of heat outflow, and the solidification time may also be close to or shorter than Earth's \citep{foley2024exoplanet}.
The amount of water considered in the system also affects the solidification rate by affecting the outgoing flux from the surface of the planet.
More water vapor in the atmosphere results in greenhouse effect which traps more heat, delaying the solidification. 
\citet{barth2021magma} compared the solidification time for TRAPPIST-1 planets with different amounts of water in the interior, which reflects in the atmospheric water vapor.
Our simulations for GJ 486b show that reduction in initial amount of water in the magma ocean by a factor of 10 causes the solidification time to reduce by a factor of 10. The effects of changing initial water on the thermal evolution of GJ 486b have been shown in Appendix \ref{apdx:sec:water_tests}.
Larger planets contain more water, leading to more water vapor in their atmospheres which causes extended mantle solidification. 
The Earth has undergone several impacts in the early stage of its formation; our Moon formation is one such event \citep{canup2004dynamics, nakajima2015melting}.
These impacts provide tremendous amounts of energy which is sufficient to melt a huge part of the mantle and prolong its solidification \citep{sleep2014terrestrial}, eroding away a large portion of atmosphere at the same time. 
All these processes and their interplay potentially affect the planet's magnetic field generation and tectonic activity too.
Furthermore, the elevated pressures and temperatures could influence mineral phase transitions, contributing to a more complex interior structure, especially towards the deep interior. 


The quantity of water plays a crucial role in the thermal evolution of a planet. 
Deciphering the initial water content on a planet is very challenging. Previous studies by \citet{nuth2008volatile, marty2012origins} discuss the major volatile concentration and their deposition on Earth. The Earth is speculated to have more water delivered by planetesimals during its formation \citep{morbidelli2000source}. However, the amount of water on early Earth is still in debate. \citet{dorn2021hidden} investigated that magma ocean planets can hold several Earth Oceans (EO) of water. 
Simulations by \citet{bower2022retention} show the possibility of higher amounts of water retained in the magma oceans in higher C/H ratio atmospheres.
Our simulations for 10 EO initial water show that the GJ 486b retains a huge amount of water in its atmosphere at the end of the simulation, which agree with the water-rich atmosphere observations of \citet{moran2023high}. 
The gradual buildup of atmospheric water and oxygen results from the limited XUV radiation available for dissociation and escape of the water molecules.
The pressure buildup in the atmosphere reaches a steady state after the planet solidifies in about 3 Myrs; after this, the changes are not significantly visible on the plots.

The interplay between outgassing, photolysis, and atmospheric escape shapes the planet's volatile inventory, ultimately influencing its long-term climate and atmospheric composition.
These processes could determine whether the planet retains a substantial atmosphere, potentially impacting surface conditions such as temperature and pressure. 
For example, Earth-like planets with a sustained release of volatiles may maintain atmospheres capable of supporting liquid water, while others may lose their atmospheres to space, resulting in barren, Mars-like conditions. 
In this work, the loss of atmosphere is primarily due to XUV-driven photodissociation and hydrodynamic escape. However, there are several other mechanisms that can lead to loss of atmosphere. 
Photodissociation and hydrodynamic escape can still happen after the XUV radiation phase of the planet has ended. This is because UV (not necessarily XUV) can also dissociate water, similar to XUV dissociation discussed in the paper \citep{guo2019effect, munoz2023heating, tabone2024oh}. Additionally, thermal escape, where high temperatures give molecules enough kinetic energy to escape the planet's gravity, can also contribute to water loss \citep{tian2009thermal, johnson2010thermally, koskinen2014thermal}. Solar wind stripping, particularly for planets without a strong magnetic field and around active stars, can erode the atmosphere over time, especially during periods of high solar activity \citep{rodriguez2019erosion, owen2019atmospheric, harbach2021stellar}. Impact erosion from large asteroid or comet impacts can strip away significant portions of the atmosphere, including water vapor \citep{kegerreis2020atmospheric}.

Our simplified treatment for the atmosphere does not differentiate between water-vapor atmosphere and formation of oceans on the planet.
Such high pressures of water in the atmosphere may lead to a reservoir of liquid water on the surface of the planet, causing a huge decrease in the atmospheric water pressure. However, in order to include the formation of liquid water oceans, a line-by-line climate model would be necessary, which is out of the scope of this paper. While our results suggest a water-rich atmosphere, similar to the water-rich scenario constrained by JWST observations from \citet{moran2023high}, the current JWST data cannot definitively distinguish between a water-rich planetary atmosphere and stellar contamination. Future high-sensitivity observations, with improved wavelength coverage and resolution enabling the detection of trace gases in the atmosphere, may help resolve this ambiguity.

\subsection{Limitations of the model and future outlook} \label{subsec:limitations} 


Both the structure and evolution models involve several challenges, which are addressed in our model through simplifications, parameterizations, and assumptions. Our structure model is based on a simplified version of Earth's composition; however, in reality, Earth's composition is a complex mix of various minerals and rocks, distributed non-uniformly across different layers. Changes in mineral forms at different depths can occur and may be highly sensitive to variations in temperature and pressure. 
Determining the exact composition of various types of rocky exoplanets is challenging, primarily due to their diversity and limited experimental and numerical data on material properties. This challenge is compounded when dealing with properties of existing materials at the extreme pressures found in planetary interiors, such as the core-mantle boundary pressure of GJ 486b, which is comparable to the pressure at Earth's core. Understanding these variations highlights the diversity of rocky exoplanets and emphasizes the need for models that account for differences in mass and composition. 
Studies such as \citet{stixrude2005structure, mosenfelder2009mgsio3, sakai2016experimental, spaargaren2023plausible} have played a pivotal role in expanding these knowledge barriers.

The evolution model also has a few limitations. We have considered a simple 1-D convective model for our interior, whereas in reality, the mantle convects in three dimensions at varying rates. This could lead to inhomogeneities within the mantle, resulting in a varied composition. 
One such example is the case of tidally locked planets, which can exhibit hemispherical differences in mantle convection and outgassing \citep{van2011mantle, meier2021hemispheric}. \citet{meier2024geodynamics} investigated the possibility of differing hemispheric convection patterns on GJ 486b. The planet is expected to show different surface temperatures, and hence different rates of mantle cooling and outgassing. As a result, atmospheric escape would be faster on the dayside than on the nightside. A nightside-to-dayside wind flow could be anticipated in 3D simulations of this planet.
In our model. several properties, such as mineral compositions, melt-solid partition coefficients, and mantle density, have been assumed to be independent of pressure, temperature, and radial distance. In practice, these properties are dependent on both temperature and pressure and vary with depth.

While our model currently assumes the presence of only one volatile, such as water, the inclusion of additional volatiles and their coupled interior-atmosphere evolution in future work may provide further constraints from upcoming JWST observations. 
Validating the evolutionary timescale remains another significant challenge. Many properties in our model are not well constrained, even for Earth—the planet we know best—making it highly challenging to draw conclusions about distant planets whose interiors are impossible to observe.

A key limitation of our model is its inability to comment on the tectonic behavior of the planet. Earth has a system of tectonic plates that has played a crucial role in the evolution of the upper mantle and the formation of its complex crust \citep{condie2013plate, hawkesworth2016tectonics, hawkesworth2020evolution}. 
Deciphering about a planet’s tectonics regime is extremely difficult, which requires a combination of modeling and observational data from the surface of the planet. While Mercury and Mars are believed to have stagnant lids, the tectonics regime of Venus is still not clear. \citet{lourencco2023past} have shown a detailed discussion on the possibility of different types of tectonics on rocky planets, including the solar system planets. \citet{valencia2007inevitability, foley2012conditions} highlighted the possibility of plate tectonics on super-Earths. \citet{foley2014initiation} investigated the mechanism of origin of plate tectonics on Hadean Earth and speculated similar origin of tectonics on super-Earths. Other tectonics regimes have also been speculated for super-Earth planets \citep{o2007geological, tackley2023tectono}. However, these studies require more complex modeling, which is beyond the scope of the present work.


\section{Conclusions} \label{sec:conclusion}

Our study utilizes a 1D self-consistent coupled interior structure and evolution model, \texttt{SERPINT}, to simulate the interior, thermal, and atmospheric evolution of the rocky exoplanet GJ 486b. The following key points summarize our findings:


\begin{itemize}
    
    \item The structural model, validated against the Preliminary Reference Earth Model (PREM), closely matches Earth's core-radius fraction and density profile, with minor deviations due to the assumption of a fully liquid iron core, whereas PREM accounts for Earth's solid inner core and liquid outer core.
    
    \item Our model predicts Earth's core-radius fraction to be approximately 0.538 $R_{\oplus}$, slightly lower than the PREM value of 0.546 $R_{\oplus}$, highlighting the simplifications inherent in our model.

    \item For GJ 486b, the model predicts a core-radius fraction of 0.537 $R_{\text{p}}$, with a core radius approximately 1.34 times larger than Earth's, reflecting its greater mass and similar internal structure.

    \item The density at the center of GJ 486b is predicted to be 17.57 kg/m$^3$, significantly higher than Earth's, due to the planet's higher mass. This is attributed to a core pressure of 1171 GPa, which is three times that of Earth's.

    \item The thermal evolution model predicts that GJ 486b's mantle, starting at a potential temperature of 5000 K, cools and solidifies over approximately 0.93 million years.

    \item The extended solidification of the upper mantle compared to the lower mantle is attributed to its higher solidus temperature, significantly affecting the overall cooling rate.

    \item The cooling rate decelerates as solidification proceeds, due to latent heat release during crystallization, with the solidification radius reaching the upper-lower mantle boundary at 2600 K.


    \item As GJ 486b cools, water initially dissolved in the magma is progressively released into the atmosphere, increasing atmospheric water pressure and leading to dissociation into hydrogen and oxygen.

    \item Our models show the presence of a thick water atmosphere which is consistent with the observations of \citet{moran2023high}.

    \item The atmospheric pressure of water increases from 61.5 bars to around 1275 bars during the early stages of solidification, followed by a steady-state of 1062 bars as the planet solidifies completely.

    \item The model shows a gradual buildup of atmospheric oxygen due to limited XUV radiation available for further dissociation and escape, with oxygen pressure stabilizing as XUV radiation decreases.

    \item The interplay between outgassing, photolysis, and atmospheric escape shapes the planet's volatile reservoirs, impacting its long-term atmospheric composition.
\end{itemize}

\section*{Acknowledgments}
L.M. acknowledges the financial support of the DAE and DST-SERB research grants (MTR/2021/000864) from the Government of India for supporting this work. L.M. also extends thanks to Priyadarshi Chowdhury for the insightful discussions at SEPS, NISER, related to the geological evolution of the Earth. We would like to thank the anonymous referee for constructive comments that helped to improve the manuscript.
 


\newpage 
\appendix

\setcounter{section}{0} 
\renewcommand{\thesection}{\Alph{section}} 
\refstepcounter{section}

\section*{APPENDIX A: core impurities}
\label{apdx:sec:impurity} 


The Earth's core is primarily composed of liquid $\varepsilon$-Fe. We utilize the Holzapfel equation of state (EOS) due to its closed form, resulting in significantly fewer volume residuals for pressures above 234 GPa compared to the Vinet and BME3 EOS \citep{hakim2018new}. The calculated EOS for liquid $\varepsilon$-Fe suggests a good fit for the Earth’s core density profile. While the Vinet and BME3 EOS are general forms used to describe the behavior of various materials under high-pressure conditions, they are often applied to pure iron in studies of the Earth’s core. 

However, the Earth’s iron core contains various impurities that reduce its density. We use the Holzapfel EOS to model the Earth’s core density profile more accurately. Additionally, we incorporate a weighted average of the respective impurities to account for the observed density decrement. A weighted average of impurity is given by
\begin{equation}
w_{\mathrm{im}}=\frac{\mathrm{Si}_{\mathrm{im}} w_{\mathrm{Si}}+\mathrm{O}_{\mathrm{im}} w_{\mathrm{O}}+\mathrm{S}_{\mathrm{im}} w_{\mathrm{S}}}{\text { total impurity }} 
\end{equation}
where $\mathrm{X}_{\mathrm{im}}$ and $w_{\mathrm{X}}$ are the added impurity percentages and molar mass of species X, respectively. The total impurity is given by
\begin{equation}
\text{total impurity} = \mathrm{Si}_{\mathrm{im}}+\mathrm{O}_{\mathrm{im}}+\mathrm{S}_{\mathrm{im}}    
\end{equation}

We took the values $w_{\mathrm{Fe}}=55.845$ g mol$^{-1}$, $w_{\mathrm{Si}}=28.0855$ g mol$^{-1}$,, $w_{\mathrm{S}}=32.065$ g mol$^{-1}$, and $w_{\mathrm{O}}=15.999$ g mol$^{-1}$,.
The average molar mass of the impure iron core is then defined as
\begin{equation}
\mathrm{Fe}_{\mathrm{im}}=(1-\text { total impurity }) \times w_{\mathrm{Fe}}+\text { total impurity } \times w_{\mathrm{im}}
\end{equation}
Hence, the impurity factor is given by
\begin{equation}
f=1-\frac{\mathrm{Fe}_{\mathrm{im}}}{w_{\mathrm{Fe}}}
\end{equation}

Hence, the modified Holzapfel EOS for the core becomes
\begin{equation}
 P = P_0 + 3 K_{0} \xi'^{-5/3}\left[1+c_2 \xi'^{1/3}\left(1-\xi'^{1/3}\right)\right] \times \left[1-\xi'^{1/3}\right] \exp \left[c_0\left(1-\xi'^{1/3}\right)\right]
\end{equation}
where $\xi' = \dfrac{\rho}{\rho_0(1-f)}$ (refer Eqn. \ref{eqn:holzapfel} for the definition of other terms).
This modification has enabled the addition of lighter elements to the core without the need to change the EOS.

\refstepcounter{section}

\section*{APPENDIX B: Atmospheric Escape} 
\label{apdx:sec:escape} 

High-energy ultraviolet photons photolyze water molecules into hydrogen and oxygen atoms \citep{hunten1973escape, hunten1987mass}. We assume that all hydrogen and oxygen originate from water photolysis. Since hydrogen has a lower atomic mass, it escapes into space more easily through mechanisms, such as Jean’s escape or hydrodynamic escape. This results in a loss of water mass, which can be quantified by summing the masses of the escaping hydrogen and oxygen \citep{luger2015extreme}.
\begin{equation}
    \dot{M}_{EL} = \dot{m}_{\text{H}}^{\uparrow} + \dot{m}_{\text{O}}^{\uparrow}
    \label{eqn:M_dot_EL_2}
\end{equation}

The equations presented describe the steady-state conditions where the escape fluxes of hydrogen (\(\dot{m}_{\text{H}}^{\uparrow}\)) and oxygen (\(\dot{m}_{\text{O}}^{\uparrow}\)) are balanced by their production rates from water photolysis.
Here $\dot{M}_{EL}$ denotes the total mass loss rate of escaping gases from the planet's atmosphere.
Based on the masses of H and O, and number of atoms in each molecule of water, we can relate the production rate of hydrogen and oxygen as  \citep{luger2015extreme}
\begin{equation}
    \dot{m}_{\text{O}} = 8\dot{m}_{\text{H}}
    \label{eqn:m_H_and_O_dot}
\end{equation}
Partitioning the escaping oxygen and the oxygen being accumulated in the atmosphere gives 
\begin{equation}
    \dot{m}_{\text{O}}^{\uparrow} + \dot{m}_{\text{O}}^{\text{atm}} = 8 \dot{m}_{\text{H}}^{\uparrow}
    \label{eqn:m_O_dot_up_and_atm}
\end{equation}
Solving these equations results in the final form 
\begin{align}
    \dot{m}_{\text{H}}^{\uparrow} = \left( \dfrac{\Phi_{\text{H}}}{\Phi_{\text{H}} + 16 \Phi_{\text{O}}} \right) \dot{M}_{EL}
\end{align}
Here $\Phi_{\text{H}}$ and $\Phi_{\text{O}}$ are the photolysis rates of hydrogen and oxygen, respectively. The ratio $\dfrac{\Phi_{\text{H}}}{\Phi_{\text{H}} + 16 \Phi_{\text{O}}}$ determines the fraction of total mass loss attributed to hydrogen escape.

\begin{align}
    \dot{m}_{\text{O}}^{\uparrow} = \left( \dfrac{16 \Phi_{\text{O}}}{\Phi_{\text{H}} + 16 \Phi_{\text{O}}} \right) \dot{M}_{EL}
\end{align}
\begin{align}
    \dot{m}_{\text{O}}^{\text{atm}} = 8 \left( \dfrac{\Phi_{\text{H}} -2 \Phi_{\text{O}}}{\Phi_{\text{H}} + 16 \Phi_{\text{O}}} \right) \dot{M}_{EL}
\end{align}

The differential escape of hydrogen and oxygen can lead to significant changes in the atmospheric composition over geological timescales. For instance, a planet initially rich in water may experience substantial hydrogen loss, potentially resulting in an atmosphere enriched in oxygen. By quantifying the escape rates of hydrogen and oxygen, we gain insights into the long-term stability of water on a planet and the potential for developing an oxygen-rich atmosphere.

\refstepcounter{section}

\section*{APPENDIX C: Effect of initial amount of water on thermal evolution}
\label{apdx:sec:water_tests}

The quantity of water plays a crucial role in the thermal evolution of a planet. As detailed in Section \ref{subsubsec:volatile}, water is released from the planet's interior during the solidification process, contributing to the formation of a secondary water vapor atmosphere. An increase in atmospheric water vapor enhances the greenhouse effect, which in turn traps more heat and delays the solidification process.

\begin{figure*}[ht]
\centering
\includegraphics[width=0.82\textwidth]{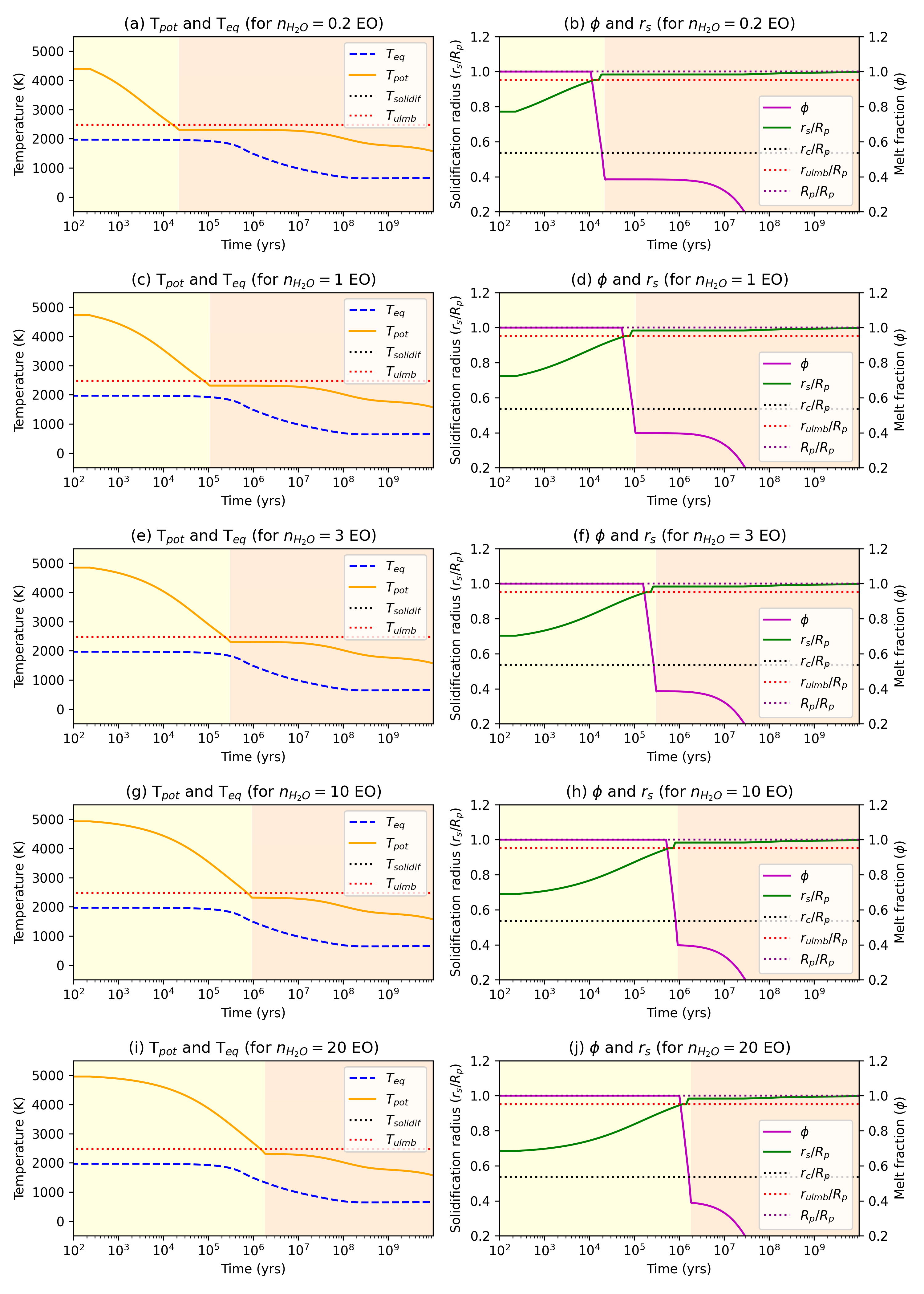}
\caption{Evolution of thermal properties of GJ 486b with 0.2 EO, 1 EO, 3 EO, 10 EO and 20 EO of initial water. The plots in the left column represent the mantle potential temperature evolution, while the plots in the right column illustrate the evolution of solidification radius fraction and melt fraction. The dotted lines denote the core radius ($r_c$), upper-lower mantle boundary radius ($r_{\text{ulmb}}$) and the planet radius ($R_p$), all normalized to planet radius.}
\label{fig:GJ486b_thermal_H2O_grid}
\end{figure*}

Deciphering the initial water content on a planet is challenging. \citet{dorn2021hidden} investigated that magma ocean planets can hold several Earth Oceans (EO) of water. Additionally, the cores of planets can also contain substantial reservoirs of water \citep{li2020earth}. We examined the effect of varying initial water content on the thermal evolution of GJ 486b. Our simulations encompassed five scenarios: 0.2 EO, 1 EO, 3 EO, 10 EO, and 20 EO, representing a range from very low to high water content.
Fig. \ref{fig:GJ486b_thermal_H2O_grid} shows the thermal evolution and evolution of solidification radius and melt fraction for GJ 486b for the five cases.
Our results indicate that variations in initial water content lead to significant differences in the solidification time of the planet. Increasing the water content delays the planet's solidification by several thousand years.

Fig. \ref{fig:GJ486b_thermal_H2O_grid}(a) illustrates the thermal evolution of GJ 486b with 0.2 Earth Oceans (EO) of initial water, representing a relatively dry scenario. We observe that the planet solidifies as early as 0.02 million years (Myrs), as shown in Fig. \ref{fig:GJ486b_thermal_H2O_grid}(b). Initially, the planet cools rapidly until solidification, after which it transitions to a slower cooling rate in the solid-state evolution regime. Other aspects of the evolution follow patterns similar to those discussed in Section \ref{subsec:thermal}. 
For the case of 1 EO of initial water, the solidification process is delayed by 0.09 Myrs; the planet rapidly cools before transitioning to the solid-state evolution regime at 0.11 Myrs, as shown in Fig. \ref{fig:GJ486b_thermal_H2O_grid}(c) and \ref{fig:GJ486b_thermal_H2O_grid}(d). Similarly, for the cases of 3 EO, 10 EO, and 20 EO, the planet solidifies after 0.3 Myrs, 0.93 Myrs, and 1.82 Myrs, respectively, as shown in Fig. \ref{fig:GJ486b_thermal_H2O_grid}. Further details on the evolution of other properties can be found in Section \ref{subsec:thermal}.

It was also observed that as the planet enters the solid-state evolution regime, it cools very slowly, resulting in nearly identical mantle potential temperatures at the conclusion of all simulations. This indicates that the amount of water does not impact the mantle potential temperature of the planet after a few billion years of evolution.

\newpage
\bibliography{references}{}
\bibliographystyle{aasjournal}



\end{document}